%% file: main.tex
\documentclass[%
 aip,
amsmath,amssymb,
 reprint,%
nofootinbib
]{revtex4-1}

\usepackage{graphicx}
\usepackage{dcolumn}
\usepackage{bm}
\usepackage{url}
\usepackage[utf8]{inputenc}
\usepackage[T1]{fontenc}
\usepackage{etoolbox}
\usepackage[nolist]{acronym}
\usepackage[inline]{enumitem}
\usepackage{booktabs}
\usepackage{xspace}
\usepackage{amsmath,amssymb}
\usepackage{xcolor}
\usepackage{xfrac}

\input{dmradio_acronyms}

\newcommand{\cs}[1]{\textcolor{magenta}{CS: #1}}

\newcommand{\dVdPhi}{$\left. \left( \partial V_{\rm out}/\partial \Phi \right) \right|_{I_{\rm out}}$}
\newcommand{\dIdPhi}{$\left. \left( \partial I_{\rm out}/\partial \Phi \right) \right|_{V_{\rm out}}$}
\newcommand{\dVdI}{$\left. \left( \partial V_{\rm out}/\partial I_{\rm out} \right) \right|_{\Phi}$}

\newcommand{\Lsq}[1]{$L_{\rm sq}$}
\newcommand{\Lin}[1]{$L_{\rm in}$}
\newcommand{\Rj}[1]{$R_{\rm j}$}
\newcommand{\kb}[1]{$k_{\rm B}$}
\newcommand{\Tj}[1]{$T_{\rm j}$}
\newcommand{\Tn}[1]{$T_{\rm n}$}
\newcommand{\Tmin}[1]{$T_{\rm min}$}
\newcommand{\CMT}[1]{CITLF2}

\begin{document}

\preprint{AIP/123-QED}

\title{Noise limits for dc SQUID readout of high-$Q$ resonators below 300 MHz}
\input{authorlist2024}

\date{\today}

\begin{abstract}
We present the limits on noise for the readout of cryogenic high-$Q$ resonators using dc Superconducting Quantum Interference Devices (SQUIDs) below 300\,MHz. 
This analysis uses realized first-stage SQUIDs (previously published), whose performance is well described by Tesche-Clarke (TC) theory, coupled directly to the resonators. We also present data from a prototype second-stage \dcsq{} array designed to couple to this first-stage SQUID as a follow-on amplifier with high system bandwidth. This analysis is the first full consideration of \dcsq{} noise performance referred to a high-$Q$ resonator over this frequency range, and is presented relative to the standard quantum limit.  We include imprecision, backaction, and backaction-imprecision noise correlations from TC theory, the noise contributed by the second-stage SQUIDs, wiring, and preamplifiers, and optimizations for both on-resonance measurements and off-resonance scan sensitivity. This architecture has modern relevance due to the increased interest in axion searches and the requirements of the DMRadio-m$^3$ axion search, which will use \dcsq{}s in this frequency range.

\end{abstract}

\maketitle

\include{sections}

\bibliography{dmradio_bibliography}

\appendix
\include{appendices}

\end{document}

%% file: dmradio_acronyms.tex
\newcommand{\DMR}{\mbox{DMRadio}\xspace}

\newcommand{\DMRm}{\mbox{\DMR-m$^3$}\xspace}

\newcommand{\dcsq}{\mbox{dc SQUID}\xspace}

\begin{acronym}
\acro{SM}{Standard Model}
\acro{QED}{quantum electrodynamics}
\acro{QCD}{quantum chromodynamics}
\acro{BSM}{beyond the standard model}
\acro{DM}{dark matter}
\acro{CDM}{cold dark matter}
\acro{GUT}{grand unification theory}
\acro{WIMP}{weakly interacting massive particle}
\acro{SHM}{Standard Halo Model}
\acro{ppm}{part-per-million}
\acro{ppb}{part-per-billion}

\acro{ADM}{axion dark matter}
\acro{ALP}{axion-like particle}
\acro{PQ}{Peccei-Quinn}
\acro{PQWW}{Peccei-Quinn-Wilczek-Weinberg}
\acro{KSVZ}{Kim-Shifman–Vainshtein–Zakharov}
\acro{DFSZ}{Dine–Fischler–Srednicki–Zhitnitsky}
\acro{LSW}{light shining through wall}

\acro{DP}{dark photon}

\acro{DR}{dilution refrigerator}
\acro{PT}{pulse tube}
\acro{OFHC}{oxygen-free, high-conductivity}

\acro{TE}{transverse electric}
\acro{TM}{transverse magnetic}
\acro{TEM}{transverse electromagnetic}
\acro{MQS}{magneto-quasistatic}
\acro{SQL}{standard quantum limit}
\acro{QND}{quantum non-demolition}

\acro{DFT}{discrete Fourier transform}
\acro{FFT}{fast Fourier transform}
\acro{SNR}{signal-to-noise ratio}
\acro{PSD}{power spectral density}
\acro{FOM}{figure of merit}

\acro{SQUID}{superconducting quantum interference device}
\acro{SQL}{standard quantum limit}

\end{acronym}

\acrodefplural{PSD}{power spectral densities}
\acrodefplural{GUT}{grand unification theories}
\acrodefplural{ppm}{parts-per-million}
\acrodefplural{ppb}{parts-per-billion}


%% file: authorlist2024.tex
\author{V.~Ankel}
\affiliation{Department of Physics, Stanford University, Stanford, CA 94305}
\affiliation{Kavli Institute for Particle Astrophysics and Cosmology, Stanford University, Stanford, CA 94305}

\author{C.~Bartram}
\affiliation{Kavli Institute for Particle Astrophysics and Cosmology, Stanford University, Stanford, CA 94305}
\affiliation{Stanford Linear Accelerator Center, Menlo Park, CA 94025}

\author{J.~Begin}
\affiliation{Department of Physics, Princeton University, Princeton, NJ 08544}

\author{C.~Bell}
\affiliation{Department of Physics, Stanford University, Stanford, CA 94305}
\affiliation{Kavli Institute for Particle Astrophysics and Cosmology, Stanford University, Stanford, CA 94305}

\author{L.~Brouwer}
\affiliation{Accelerator Technology and Applied Physics Division, Lawrence Berkeley National Laboratory, Berkeley, CA 94720}

\author{S.~Chaudhuri}
\affiliation{Department of Physics, Princeton University, Princeton, NJ 08544}

\author{John~Clarke}
\affiliation{Department of Physics, University of California, Berkeley, CA 94720}

\author{H.-M.~Cho}
\affiliation{Kavli Institute for Particle Astrophysics and Cosmology, Stanford University, Stanford, CA 94305}
\affiliation{Stanford Linear Accelerator Center, Menlo Park, CA 94025}

\author{J.~Corbin}
\affiliation{Department of Physics, Stanford University, Stanford, CA 94305}
\affiliation{Kavli Institute for Particle Astrophysics and Cosmology, Stanford University, Stanford, CA 94305}

\author{W.~Craddock}
\affiliation{Stanford Linear Accelerator Center, Menlo Park, CA 94025}

\author{S.~Cuadra}
\affiliation{Laboratory of Nuclear Science, Massachusetts Institute of Technology, Cambridge, MA 02139}

\author{A.~Droster}
\affiliation{Department of Nuclear Engineering, University of California, Berkeley, CA 94720}

\author{M.~Durkin}
\affiliation{National Institute of Standards and Technology, Boulder, CO 80305}
\affiliation{University of Colorado, Boulder, Boulder, CO 80305}

\author{J.~Echevers}
\affiliation{Department of Nuclear Engineering, University of California, Berkeley, CA 94720}

\author{J.~T.~Fry}
\affiliation{Laboratory of Nuclear Science, Massachusetts Institute of Technology, Cambridge, MA 02139}

\author{G.~Hilton}
\affiliation{National Institute of Standards and Technology, Boulder, CO 80305}

\author{K.~D.~Irwin}
\affiliation{Department of Physics, Stanford University, Stanford, CA 94305}
\affiliation{Kavli Institute for Particle Astrophysics and Cosmology, Stanford University, Stanford, CA 94305}
\affiliation{Stanford Linear Accelerator Center, Menlo Park, CA 94025}

\author{A.~Keller}
\affiliation{Department of Nuclear Engineering, University of California, Berkeley, CA 94720}

\author{R.~Kolevatov}
\affiliation{Department of Physics, Princeton University, Princeton, NJ 08544}

\author{A.~Kunder}
\affiliation{Department of Physics, Stanford University, Stanford, CA 94305}
\affiliation{Kavli Institute for Particle Astrophysics and Cosmology, Stanford University, Stanford, CA 94305}

\author{D.~Li}
\affiliation{Stanford Linear Accelerator Center, Menlo Park, CA 94025}

\author{N.~Otto}
\affiliation{Department of Physics, Princeton University, Princeton, NJ 08544}

\author{K.~M.~W.~Pappas}
\affiliation{Laboratory of Nuclear Science, Massachusetts Institute of Technology, Cambridge, MA 02139}

\author{N.~M.~Rapidis}
\affiliation{Department of Physics, Stanford University, Stanford, CA 94305}
\affiliation{Kavli Institute for Particle Astrophysics and Cosmology, Stanford University, Stanford, CA 94305}

\author{C.~P.~Salemi}
\email{salemi@berkeley.edu}
\affiliation{Department of Physics, Stanford University, Stanford, CA 94305}
\affiliation{Kavli Institute for Particle Astrophysics and Cosmology, Stanford University, Stanford, CA 94305}
\affiliation{Stanford Linear Accelerator Center, Menlo Park, CA 94025}

\author{D.~Schmidt}
\affiliation{National Institute of Standards and Technology, Boulder, CO 80305}

\author{M.~Simanovskaia}
\affiliation{Department of Physics, Stanford University, Stanford, CA 94305}
\affiliation{Kavli Institute for Particle Astrophysics and Cosmology, Stanford University, Stanford, CA 94305}

\author{J.~Singh}
\affiliation{Department of Physics, Stanford University, Stanford, CA 94305}
\affiliation{Kavli Institute for Particle Astrophysics and Cosmology, Stanford University, Stanford, CA 94305}

\author{P.~Stark}
\affiliation{Department of Physics, Stanford University, Stanford, CA 94305}
\affiliation{Kavli Institute for Particle Astrophysics and Cosmology, Stanford University, Stanford, CA 94305}
\affiliation{Stanford Linear Accelerator Center, Menlo Park, CA 94025}

\author{C.~D.~Tesche}
\affiliation{Department of Psychology, University of New Mexico, Albuquerque, NM  87131-1161, USA}

\author{J.~Ullom}
\affiliation{National Institute of Standards and Technology, Boulder, CO 80305}

\author{L.~Vale}
\affiliation{National Institute of Standards and Technology, Boulder, CO 80305}

\author{E.~C.~van~Assendelft}
\affiliation{Department of Physics, Stanford University, Stanford, CA 94305}
\affiliation{Kavli Institute for Particle Astrophysics and Cosmology, Stanford University, Stanford, CA 94305}

\author{K.~van~Bibber}
\affiliation{Department of Nuclear Engineering, University of California, Berkeley, CA 94720}
\affiliation{Physics Division, Lawrence Berkeley National Laboratory, Berkeley, CA 94720}

\author{M.~Vissers}
\affiliation{National Institute of Standards and Technology, Boulder, CO 80305}

\author{K.~Wells}
\affiliation{Department of Physics, Stanford University, Stanford, CA 94305}

\author{J.~Wiedemann}
\affiliation{Department of Physics, Princeton University, Princeton, NJ 08544}

\author{L.~Winslow}
\affiliation{Laboratory of Nuclear Science, Massachusetts Institute of Technology, Cambridge, MA 02139}

\author{D.~Wright}
\affiliation{Kavli Institute for Particle Astrophysics and Cosmology, Stanford University, Stanford, CA 94305}
\affiliation{Department of Electrical Engineering, Stanford University, Stanford, CA 94305}

\author{A.~K.~Yi}
\affiliation{Kavli Institute for Particle Astrophysics and Cosmology, Stanford University, Stanford, CA 94305}
\affiliation{Stanford Linear Accelerator Center, Menlo Park, CA 94025}

\author{B.~A.~Young}
\affiliation{Department of Physics, Santa Clara University, Santa Clara, CA 95053}


%% file: sections.tex
\section{Introduction}
\label{sec:intro}

SQUIDs have been used as highly sensitive magnetic flux and current amplifiers for sixty years \cite{jaklevic1964quantum,clarke2004squid};  \dcsq{}s are well established in applications ranging from fundamental physics to prospecting and medical imaging. In this paper we focus on their use to read out high quality factor ($Q$), low-temperature resonators at very high frequencies (VHF) and below ($f<300$\,MHz). Resonant applications  include Weber bars for gravitational wave detection \cite{doi:10.1063/1.3002321}, low-field nuclear magnetic resonance (NMR)\cite{sleator1987nuclear}, magnetic resonance imaging (MRI)\cite{hilbert1985dc} measurements, and searches for light-dark-matter candidates including quantum chromodynamics (QCD) axions and hidden photons \cite{Brouwer:2022DMRm,alshirawi2023electromagnetic,PhysRevLett.124.241101,Gramolin2020a,PhysRevLett.126.041301,chakrabarty2023low}. The sensitivity of these applications and experiments directly depends on the noise of the SQUID readout chain.

Weak coupling between a high-$Q$, low-temperature resonator and the first-stage \dcsq{} can be optimal. In this weak-coupling regime, optimized \dcsq{} performance is consistent with the seminal work by Tesche and Clarke \cite{tesche1977dc,tesche1979dc,clarke1979optimization} (hereafter referred to as TC). This theoretical analysis established the conditions to optimize the single-stage \dcsq{} noise performance and allows us to calculate the imprecision and backaction noise, as well as their correlations, at optimized conditions.  The imprecision noise is defined as the noise when no input circuit is attached and the input terminals are open, in which case there can be no backaction-current response of the input circuit (and hence no backaction noise contribution on the output). When coupled to an input circuit, the backaction noise term presents as a voltage applied to the input circuit, driving an input current signal that is partially correlated with the imprecision current noise in the SQUID.  We typically refer all noise contributions to the SQUID input.

\begin{figure}[t]
    \centering
    \includegraphics[width=\linewidth]{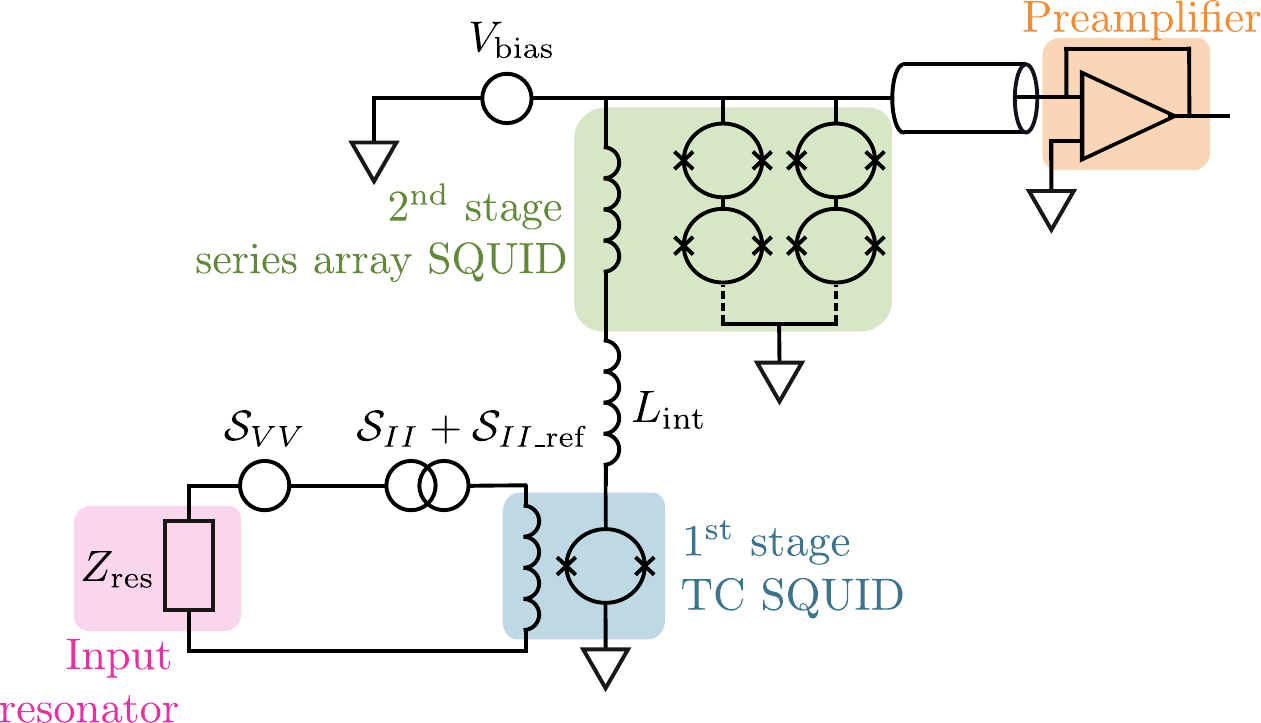}
    \caption{The full SQUID amplifier chain coupled to an input resonator with complex impedance $Z_{\rm res}$ via input inductor $L_{\rm in}$. $S_{\rm VV}$ is the PSD of the backaction voltage noise acting on the first-stage input.  $S_{\rm II}$ is the PSD of the 1st-stage \dcsq{} imprecision noise referred to a current in the input, and $S_{\rm II\_ref}$ is the imprecision noise of the rest of the amplifier chain referred to the 1st stage input. The 1st- and 2nd-stage SQUIDs are coupled via a small interconnect inductor, $L_{\rm int}$. The first-stage SQUID is voltage biased in series with the input coil of the 2nd-stage series-array \dcsq{}. The output of the 2nd-stage array is fed into a warm or cryogenic preamplifier with a 50\,$\Omega$ input inductance via a 50\,$\Omega$-impedance cable.}
    \label{fig:SQUIDChain}
\end{figure}

This work lays out the noise limits for a SQUID-based amplifier chain coupled to a resonator (see Fig.~\ref{fig:SQUIDChain}) under the conditions of weak input coupling, high frequency bandwidth, and low dynamic range (thus relaxing the need for fast flux feedback).  The full system noise comprises contributions from the first-stage \dcsq{} as well as from second-stage SQUID circuits and downstream electronics, including those at room temperature. We present gain, impedance, and noise data measured for a prototype second-stage SQUID circuit for this application.  We also discuss several options for follow-on amplification, including commercial options at cryogenic and room temperatures as well as a custom, high-bandwidth amplifier based on previously published work.  We finally combine the previously published theoretical and experimental analysis of first-stage SQUIDs with the second-stage data and follow-on amplifier specifications to consider the full system noise that can be achieved in weakly coupled resonators.

The noise performance of this SQUID amplifier chain can be optimized for different applications.  In this work, we initially compute the on-resonance noise match conditions to minimize the \dcsq{} noise temperature and optimize the resulting on-resonance performance. We also analyze the optimal noise temperature that can be achieved by a dc SQUID coupled to an arbitrary complex input impedance. We show that this noise temperature is equivalent to that which can be achieved by a single dc SQUID coupled to an LC resonator that has been optimally detuned to minimize noise temperature, as computed by Clarke\cite{clarke1979optimization}. This result shows, for the first time, that an LC resonator can achieve the optimum noise match to a dc SQUID, albeit in a narrow bandwidth. We further extend this analysis to include the noise contributions from follow-on amplifier stages.

Applications, including wavelike dark matter searches, benefit from optimizing the noise match to maximize the sensitivity of a wide frequency scan, rather than maximizing sensitivity on resonance or in a narrow bandwidth around a particular detuned frequency. The amplifier chain described in this paper, while applicable to a broad range of resonant applications, is well suited to resonant, low-mass axion dark matter searches such as the Dark Matter Radio experiments, specifically \DMRm{}.  Axion signals can be captured in resonant circuits, but to detect this tiny signal, the resonators must be optimally coupled to low-noise amplifiers.  The expected signals have a small dynamic range; however, the signal frequency is unknown, so the amplifier must be able to operate over a wide bandwidth of resonator frequency tuning ($5$--$200\,$MHz in the case of \DMRm{})  \cite{Brouwer:2022DMRm,alshirawi2023electromagnetic}.




\section{dc SQUID Response and Noise}
\label{sec:squidnoise}

In this section, we use the Tesche and Clarke (TC) SQUID noise theory \cite{tesche1977dc,tesche1979dc,clarke1979optimization} to determine the noise properties of a first-stage \dcsq{}.  Dc SQUIDs optimized according to TC have measured white-noise levels consistent with the TC predictions, as long as they are properly biased and shielded and are weakly coupled to their input circuits (see, for example, Wellstood\cite{wellstood1987low,wellstood1988excess}). Such SQUIDs have also shown a low-frequency noise component that is not predicted by TC theory, which is not significant for the resonator frequencies considered in this work \cite{anton2013magnetic}. 

TC theory predicts the three response functions for an optimized \dcsq{}: the flux-to-voltage response \dVdPhi{}, the flux-to-current response \dIdPhi{}, and the current-to-voltage response (the dynamic resistance) \dVdI{}. Here we will use these response functions to refer the TC noise terms to the input circuit as imprecision noise current ($S_{\rm II}$), backaction noise voltage ($S_{\rm VV}$), and their correlations ($S_{\rm IV}$)\footnote{This work considers the \dcsq{} used as an ``op-amp mode" amplifier, measuring current.  This is in contrast to a ``scattering mode'' or RF amplifier, which measures a transmitted wave (see \cite{Clerk2010} for the theory or, as an example of a scattering-mode amplifier, the microstrip SQUID amplifiers used in ADMX \cite{o2020microstrip}).  For an op-amp mode current amplifier such as the \dcsq{}, all noise sources can be referred to imprecision input current noise and backaction input voltage noise terms as well as their correlations \cite{Clerk2010}. }.  The TC theory is well established, but these parameters will be needed in this form for the analysis of optimized \dcsq{} matching to HF and VHF resonant circuits in the following sections.

The noise of a full \dcsq{} chain may include contributions from the first- and second-stage \dcsq{}s, the wiring, and the room-temperature preamplifier. In principle, if the first-stage \dcsq{} is coupled sufficiently strongly to the second-stage \dcsq{} (which may be achieved at low frequency), the first-stage SQUID noise will dominate. We first consider the case of negligible follow-on amplifier chain noise and calculate the noise of the first-stage \dcsq{} alone. We proceed in Section~\ref{sec:combo} to include the noise from the second-stage SQUID and preamplifier. 

\begin{figure}[t]
    \centering
    \includegraphics[width=\linewidth]{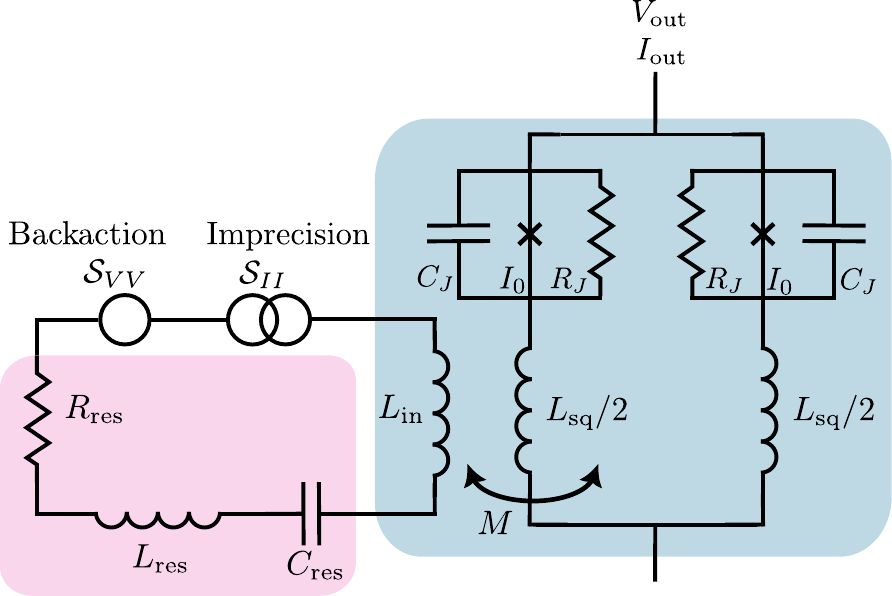}
    \caption{Circuit diagram of a TC \dcsq{} (blue highlight) coupled to a LRC series resonator (pink highlight). The \dcsq{} loop inductance \Lsq{} is divided symmetrically on both sides of the SQUID loop, coupling to the input coil \Lin{} with mutual inductance 
    $M=\kappa \sqrt{L_{\rm in} L_{\rm sq}}$, where $\kappa\in[0,1]$ is the coupling constant. $\kappa=1$ corresponds to full flux sharing between the two inductors whereas $\kappa=0$ corresponds to zero flux sharing. A variable transformer is used to tune the value of $M$ to optimize signal-to-noise ratio.
    $S_{\rm II}$ is the PSD of the \dcsq{} imprecision noise referred to a current in the input. $S_{\rm VV}$ is the PSD of the backaction voltage noise, which drives a physical noise current in the resonator.}
    \label{fig:ResonantCoupling}
\end{figure}

According to TC noise analysis, optimal operation of a single, symmetric, weakly coupled dc SQUID is under the following conditions:
\begin{equation}\begin{aligned}\label{eqn:TCassumptions}
    \beta_{\rm L} &\equiv L_{\rm sq} I_0/\Phi_0 \approx 1 \\
    \beta_{\rm C} &\equiv \pi I_0 R_{\rm j}^2 C_{\rm j}/\Phi_0 \approx 1 \\
    \Gamma &\equiv 2 \pi k_{\rm B} T_{\rm j} / I_0 \Phi_0 \lessapprox 0.05.
\end{aligned}\end{equation}
Here, $\beta_{\rm L}$ and $\beta_{\rm C}$ describe the inductive and capacitive properties of the \dcsq{} and $\Gamma$ is a unitless figure of merit for the temperature of the electrons in the \dcsq{} shunt resistance (this value is achievable for \dcsq{}s cooled in dilution refrigerators). As shown in Fig.~\ref{fig:ResonantCoupling}, \Lsq{} is the SQUID inductance (split symmetrically on each side of the SQUID loop),  $C_{\rm j}$ is the stray capacitance of a single Josephson junction, $I_0$ is the critical current of a single junction, and \Rj{} is the shunt resistance across a single junction.  \kb{} is the Boltzmann constant, $\Phi_0$ is the flux quantum, and \Tj{} is the physical temperature of the electrons in the \dcsq{} shunts.

 When the \dcsq{} parameters are optimized to follow Eq.~\ref{eqn:TCassumptions} and are operated at optimum flux bias ($\Phi=\Phi_0/4$) and current bias ($I_{\rm out}=1.8 I_0$), and at frequencies well below the Josephson frequency $f_{J}=V/\Phi_{0}$ ($V$ being the voltage across the junction), we can define the following transfer functions\footnote{Here we actually assume $\Gamma<0.025$ instead of the nominal 0.05 from Eq.~\ref{eqn:TCassumptions}.  If the nominal value is used, a slightly lower value of \dVdPhi{} is attained, as in e.g. Ref. \cite{Bruines1982}. Ref. \cite{clarke1979optimization} uses the same value as Eq.~\ref{eq:FluxToVoltage}. See App. \ref{sec:ResponseFunctions} for further discussion.}.  The flux-to-voltage transfer function for fixed bias current $I_{\rm out}$ is\cite{clarke1979optimization,Bruines1982}
\begin{equation}\label{eq:FluxToVoltage}
    \left. \frac{\partial V_{\rm out}}{\partial \Phi} \right|_{I_{\rm out}} \approx \frac{R_j}{L_{sq}} \,.
\end{equation}

The dynamic resistance of the SQUID \dVdI{} is plotted but not numerically stated in \cite{tesche1977dc}. In App. \ref{sec:ResponseFunctions} of this work, data from \cite{tesche1977dc} and \cite{Bruines1982} are numerically fit to determine that 
\begin{equation}\label{eq:DynamicResistance}
    R_{\rm dyn}=\left. \frac{\partial V_{\rm out}}{\partial I_{\rm out}} \right|_{\Phi} \approx R_j, 
\end{equation}
which is twice the resistance of the shunts across both junctions in parallel ($R_j/2$).\footnote{For very high bias currents, $I\gg I_0$, the dynamic resistance is simply the parallel resistance of the two shunt resistors, 
$R_{\rm dyn}=R_j/2$. Some sources erroneously assume that this is also the dynamic resistance at optimal TC bias. However, at optimal TC bias, the dynamic resistance is approximately twice this value. See App. \ref{sec:ResponseFunctions} for details.}

While the TC analysis was conducted with a current-biased \dcsq{}, the \dcsq{} readout chain that we present in Fig.~\ref{fig:SQUIDChain} and Section~\ref{sec:squidchips} relies on a voltage-biased first-stage \dcsq{}. It should be emphasized that the transfer functions (Eqs.~\ref{eq:FluxToVoltage} and \ref{eq:DynamicResistance}) represent the low-frequency ($<300$\,MHz) average response. The ac Josephson junction dynamics occur near the much higher Josephson frequency ($\sim10$\,GHz for typical first-stage \dcsq{}s) and its harmonics. The TC analysis is conducted with a constant current bias, and this constant current bias must be maintained at the Josephson frequencies for TC theory to be valid.\footnote{In fact, a voltage-bias on a Josephson junction at the Josephson frequencies would lead to the trivial solution of a linear temporal evolution of the phase across the junction.}

If a voltage bias is applied across an inductance in series with the \dcsq{}, then a current bias can be maintained at the Josephson frequency while the first-stage \dcsq{} is voltage-biased at VHF and HF frequencies \cite{kiviranta2001effect}. The impedance of the inductance must be large as compared to the \dcsq{} shunt resistance at the Josephson frequency and its relevant harmonics, but small enough for the necessary response bandwidth in the VHF and HF bands. The inductance of the input coil of the second-stage SQUID is typically shorted out by the coil capacitance at microwave frequencies, so the microwave inductance must be provided by a reactive element with high enough impedance at the Josephson frequency and its harmonics. Sometimes a wirebond ($L\sim1\,$nH) is sufficient for this reactance. Sufficient inductance for well-behaved voltage-biased operation of the first-stage in a two-stage \dcsq{} is routinely achieved in applications including astronomical bolometric instruments, which are routinely and robustly implemented with large \dcsq{} channel counts \cite{ahmed2014bicep3, doriese2016developments,telescope2002scuba}.
As will be discussed in Section~\ref{sec:combo}, we assume that an additional 1\,nH inductance with good microwave properties is placed in series with the first-stage \dcsq{}. 

If the condition of current-bias at the Josephson frequency and voltage-bias at HF and VHF frequencies is obtained, then we can apply Euler's chain rule to Eqs.~\ref{eq:FluxToVoltage} and \ref{eq:DynamicResistance} to derive (see App.~\ref{sec:Euler}) the final transfer function

\begin{equation}\label{eq:FluxToCurrent}
    \left. \frac{\partial I_{\rm out}}{\partial \Phi} \right|_{V_{\rm out}} \approx -\frac{1}{L_{\rm sq}} \,,
\end{equation}
the flux-to-current response of a voltage-biased first-stage \dcsq{}.

TC provides expressions for the noise of an optimized current-biased \dcsq{} with no circuit attached to its input.  The output voltage noise power spectral density (PSD) is \cite{tesche1979dc}
\begin{equation}\label{eq:TC_VV}
    \mathcal{S}_{VV{\rm out}} \approx 16k_BT_{j}R_{j} \,.
\end{equation}
This noise arises from the Johnson thermodynamic noise in the junction resistance modified by the Josephson oscillations in the active \dcsq{} circuit. Since it is computed with an open input circuit, it can be interpreted as the imprecision noise of the first-stage \dcsq{} (referred as a voltage on its output). The full system imprecision noise will also include uncorrelated imprecision noise contributions from later amplifier stages (Section~\ref{sec:combo}).

TC also computes the physical circulating current noise in the SQUID loop when it is not coupled to an input circuit. From \cite{tesche1979dc}, this calculated physical circulating current noise has PSD
\begin{equation}\label{eq:TC_II}
    \mathcal{S}_{II{\rm circ}} \approx 11k_BT_{j}/R_{j} \,.
\end{equation}
This circulating current exists even when the input circuit is open, so it sources a portion of the output imprecision noise that is in Eq.~\ref{eq:TC_VV}. $\mathcal{S}_{VV{\rm out}}$ and $\mathcal{S}_{II{\rm circ}}$ are thus partially correlated. In addition, when connected to an input circuit, $\mathcal{S}_{II{\rm circ}}$ gives rise to a backaction voltage on the input circuit that will be calculated in Eq.~\ref{eq:BackactionVoltage}.


The correlations between the output voltage noise and the circulating current noise can be described by their cross correlation \cite{tesche1979dc},
\begin{equation}\label{eq:TC_IV}
    \mathcal{S}_{I{\rm circ}V{\rm out}} \approx 12 k_B T_{j},
\end{equation}
and we note that this cross-correlation is real.

The above equations (\ref{eq:TC_VV}-\ref{eq:TC_IV}) do not take into account quantum noise, which is a small correction for the \dcsq{}s described here, as they operate above the standard quantum limit (SQL). However, we will compare results to the SQL to confirm the validity of this assumption.

We now calculate the noise when the \dcsq{} is inductively coupled to an input circuit, under the assumption that this coupling is weak enough that Eqs.~ \ref{eq:TC_VV}-\ref{eq:TC_IV} still apply, and that signals are small enough that the flux remains close to $\Phi_0/4$. 

We make no assumptions yet about the form of the input circuit other than to specify how it is coupled, with mutual inductance, $M$, between the input circuit and the \dcsq{} loop:
\begin{equation}\label{eq:inputCoupling}
    M=\kappa\sqrt{L_{\rm sq}L_{\rm in}}\,.
\end{equation}
As shown in Fig.~\ref{fig:ResonantCoupling}, here \Lsq{} is still the SQUID inductance, $L_{\rm in}$ is the inductance of the SQUID's input coil, and $\kappa\in[0,1]$ is a constant describing the efficiency of the input coupling.

We refer all the TC noise terms to an equivalent noise on the input circuit: the referred imprecision current noise $\mathcal{S}_{II}$, the physical voltage backaction noise $\mathcal{S}_{VV}$, and their correlations $\mathcal{S}_{IV}$.
 First we consider the imprecision noise of the first-stage \dcsq{}, which is referred to an output voltage in Eq.~\ref{eq:TC_VV}. Combining Eq.~\ref{eq:TC_VV} with Eqs.~\ref{eq:FluxToVoltage} and \ref{eq:inputCoupling} gives the imprecision noise referred to a current PSD in the input circuit:

\begin{equation}\label{eq:ImprecisionCurrent}
    \mathcal{S}_{II}\equiv 
    \frac{\mathcal{S}_{VV{\rm out}}}{ \left( \left. \frac{\partial V_{\rm out}}{\partial \Phi} \right|_{I_{\rm out}} \right)^2 M^2}
    \approx \frac{16 k_B T_{j} L_{\rm sq}}{\kappa^2R_jL_{\rm in}} \,.
\end{equation}


From Eqs.~\ref{eq:TC_II} and \ref{eq:inputCoupling}, the PSD of the physical flux fluctuations in the input circuit, caused by the circulating \dcsq{} current, is
\begin{equation}\label{eq:FluxCoupled}
    M^2  \mathcal{S}_{II{\rm circ}} \approx \frac{11 k_B T_{j} \kappa^2 L_{\rm in} L_{\rm sq}}{R_j}.
\end{equation}

The backaction voltage applied to the input circuit is the time derivative of the physical flux fluctuations in the input circuit, so that
\begin{equation}\label{eq:BackactionVoltage}
    \mathcal{S}_{VV}\equiv \omega^2 M^2  \mathcal{S}_{II{\rm circ}}  \approx \frac{11 k_B T_{j}  \kappa^2 L_{\rm in} L_{\rm sq} \omega^2}{R_j}.
\end{equation}

We can similarly use Eq.~\ref{eq:TC_IV} to derive the cross correlation between the input-referred imprecision noise $\mathcal{S}_{II}$ and backaction noise $\mathcal{S}_{VV}$. The voltage backaction on the input is the derivative of the coupled flux, whereas the current imprecision on the input has no time derivative, so the real TC correlation in Eq.~\ref{eq:TC_IV} produces a purely imaginary correlation. As we will later show, the square of this correlation is important for determining the minimum noise temperature when coupled to a resonator, so it is what we compute:
\begin{equation}\label{eq:InputCorrelation}
\left(\operatorname{Im}\{\mathcal{S}_{IV}\}\right)^2\approx \left(\frac{12 k_B T_{j}  L_{sq} \omega}{R_j}\right)^2.
\end{equation}

SQUID noise figures can be presented in terms of an ``uncoupled energy sensitivity'' ($\epsilon_{\rm uc}$), which refers all imprecision noise to the equivalent flux noise in the SQUID loop if it is completely uncoupled from its input (so that no backaction contributes to the output noise). For the TC analysis and the approximations above, 

\begin{equation}\label{eq:EpsilonImprecision}
    \epsilon_{\rm uc}= \frac{\mathcal{S}_{\Phi\Phi}}{2 L_{sq}} \approx \frac{8 k_B T_{j} L_{\rm sq}}{R_j}.
\end{equation}
The prefactor is sometimes listed as 9 instead of 8 (e.g. \cite{clarke1989principles}). This slightly higher value for $ \epsilon_{\rm uc}$ corresponds to a slightly less steep flux response function, which is appropriate for $\Gamma=0.05$, whereas we assume a somewhat lower $\Gamma$. The difference is not significant. 

In practical dc SQUIDs, $\epsilon_{\rm uc} > \hbar$. However, this is not a quantum limit (it does not even include backaction), but rather a practical limit from combined design constraints on junction and stray capacitance, inductance, critical current, resistance, and self-heating temperature of the resistive shunts.

The uncoupled energy sensitivity will be used in Section~\ref{sec:ScanSensitivityMatch} in computing the optimal matching of a \dcsq{} for integrated sensitivity over frequencies detuned from the resonance, rather than just on-resonance sensitivity. 

\section{Matching to a resonator}
\label{sec:NoiseMatching}

When the inefficiently coupled input transformer of the \dcsq{} is connected to a series RLC resonant circuit (See Fig.~\ref{fig:ResonantCoupling}),
the total impedance of the resonator circuit is
\begin{equation} \label{eq:ResonatorImpedance}
Z_{\rm res}(\omega) = R + i \left( \omega L_{\rm tot} - \frac{1}{\omega C_{\rm res}}\right),
\end{equation}
where $L_{\rm tot} = L_{\rm res} + L_{\rm in}$ 
is the sum of the resonator inductance and the input inductance of the \dcsq{} transformer. The resonant frequency and quality factor are thus:
\begin{align}
    \omega_0 &= \frac{1}{\sqrt{L_{\rm tot}C_{\rm res}}} \label{eq:Omega0} \\
    Q &= \frac{\omega_0 L_{\rm tot}}{R} \label{eq:Qfactor} \,.
\end{align}


\subsection{SQUID noise temperature on resonance} \label{sec:onresonance} The noise temperature \Tn{} of the \dcsq{} is determined by the imprecision, backaction, and their correlations. We first consider the special case of the noise that will be achieved by the \dcsq{} perfectly on resonance ($\omega=\omega_0$). This is relevant in cases where the resonator is driven on-resonance or is measured when not being driven.

On resonance, the imaginary component of the input circuit impedance is zero, giving $Z_{\rm res}(\omega_0)=R$. For a \dcsq{} described by TC theory,
$\operatorname{Re}\{\mathcal{S}_{IV}\}=0$. Using the dual of Eqs. 5.73 and 5.74 in \cite{Clerk2010}, with the extra factor of 2 from our two-sided PSD, we arrive at:

\begin{equation}\label{eq:OnResonanceNoiseTemperature}
    4k_BT_{\rm n}(\omega_0) R =\mathcal{S}_{VV}+R^2\mathcal{S}_{II} \,.
\end{equation}

The minimum noise temperature (``noise match'') is therefore achieved on resonance when the input circuit impedance is:
\begin{equation}\label{eq:OnResonanceNoiseImpedance}
R=\sqrt{\frac{ \mathcal{S}_{VV}}{ \mathcal{S}_{II}}} \,.
\end{equation}

From Eqs. \ref{eq:OnResonanceNoiseTemperature} and \ref{eq:OnResonanceNoiseImpedance}, the minimum noise temperature on resonance ($T_{\rm min}(\omega_0)$) is thus:
\begin{equation}\label{eq:OnResonanceMinNoiseTemperature}
k_{\rm B}T_{\rm min}(\omega_0)=\frac{1}{2} \sqrt{ \mathcal{S}_{VV}\mathcal{S}_{II}}.
\end{equation}
This quantity is independent of correlations in the \dcsq{} noise due to matching to a real impedance (as is the case on resonance). In Section~\ref{sec:NoiseTempComplex}, we extend this analysis to a complex matching impedance, and 
as will be seen in Section~\ref{sec:ScanSensitivityMatch}, the correlations are important for the off-resonance sensitivity of a \dcsq{} in applications including axion searches.

We now consider the special case where the follow-on amplifiers introduce negligible additional noise. In principle, this condition can be achieved at low frequency with strong coupling to a second-stage amplifier. (We consider the imprecision noise contribution of the follow-on amplifiers in Section~\ref{sec:squidchips}). From  Eqs. \ref{eq:ImprecisionCurrent}, \ref{eq:BackactionVoltage}, and \ref{eq:OnResonanceNoiseImpedance},

\begin{equation}\label{eq:temp}
R\approx \frac{\sqrt{11}}{4}\kappa^2 \omega_0 L_{\rm in}.
\end{equation}

We define the ``global coupling efficiency'' to the resonator:
\begin{equation}\label{eq:kappaG}
\kappa_{\rm g}^2 \equiv \kappa^2 \frac{L_{\rm in}}{L_{\rm tot}}.
\end{equation}
Here   $\kappa_{\rm g}\in[0,1]$. Perfect global efficiency ($\kappa_{\rm g}=1)$ corresponds to perfect SQUID coil coupling ($\kappa=1$) and the condition that the input coil inductance $L_{\rm in}=L_{\rm tot}$ is the only inductance in the resonator.

From Eq.~\ref{eq:temp}, the on-resonance condition for noise matching the \dcsq{} to the resonator is then:
\begin{equation}\label{eq:OnResonanceMatch}
\kappa_{\rm g}^2 \approx \frac{4}{\sqrt{11}} \frac{R}{\omega_0 L_{\rm tot}}
= \frac{4}{\sqrt{11}} \frac{1}{Q} \approx \frac{1.2}{Q}\, .
\end{equation}
This value of $\kappa_{\rm g}$ represents the optimal coupling efficiency. Any deviation from the optimal coupling constant—either higher or lower—worsens the noise temperature. The value of $\kappa_{\rm g}$ can be tuned with a variable input transformer, as is described in Section~\ref{sec:TunableTransformer}


It is evident that the optimal global coupling efficiency $\kappa_{\rm g}$ is very small for a TC-optimized \dcsq{} coupled to a high-$Q$ resonator. It is in this regime of inefficient \dcsq{} coupling that TC theory applies.  Note that due to this weak coupling, the extra damping of the resonator by the \dcsq{} (due to lossy shunt resistors) \cite{hilbert1985measurements,van2023measurements} is reduced compared to in a strong coupling configuration; thus, SQUID-induced loss can potentially be subdominant to the internal loss of the resonator. Increasing internal resonator $Q$ to arbitrarily high values by reducing the internal loss ($R$) does not make it more difficult for SQUID loss to be made negligible, because the optimal coupling ($\kappa_g$) correspondingly decreases.

If the noise match condition is achieved, the minimum noise temperature for a TC-optimized \dcsq{} on resonance is:

\begin{equation}\label{eq:TCNoiseTemperature}
k_{\rm B}T_{\rm min}\approx 2\sqrt{11}\frac{k_{\rm B}T_{\rm j}L_{\rm sq}\omega_0}{R_j}.
\end{equation}
We note that the SQL for the amplifier noise temperature is $k_{\rm B}T_{\rm min}=\hbar \omega_0/2$ (one half photon of added noise),
and that the TC theory does not explicitly include quantum noise and thus is not valid in the low-noise limit.  The ratio of the noise temperature in Eq.~\ref{eq:TCNoiseTemperature} to the SQL for amplifier-added noise is:
\begin{equation}\label{eq:OnResonanceNoiseRatio}
\frac{\sqrt{ \mathcal{S}_{VV}\mathcal{S}_{II}}}{\hbar \omega_0} 
\approx 4\sqrt{11}\frac{k_{\rm B}T_{\rm j}L_{\rm sq}}{\hbar R_j} \,.
\end{equation}
For most practical designs, $k_{\rm B}T_{\rm min}$ is far enough above the SQL that quantum corrections can be assumed to be small.
\subsection{SQUID noise temperature for complex input impedance}\label{sec:NoiseTempComplex}
If we instead couple the input to a complex impedance (as is the case for an input resonator off-resonance), correlations can either reduce or increase the noise temperature.  From the dual of Eq.~5.76 in \cite{Clerk2010}, the minimum achievable noise temperature for an optimal complex source impedance is:
\begin{equation}\label{eq:TCNoiseTemperatureComplexMatch}
 k_BT_{\rm min}=\frac{1}{2}\sqrt{\mathcal{S}_{VV}\mathcal{S}_{II} - (\operatorname{Im}\{\mathcal{S}_{IV}\})^2} + \frac{1}{2} \operatorname{Re}\{\mathcal{S}_{IV}\} \,.
\end{equation}
For a TC-optimized \dcsq{}, $\operatorname{Re}\{\mathcal{S}_{IV}\}=0$, and therefore the minimum noise temperature for matched complex input impedance is:
\begin{equation}\label{eq:TCNoiseTemperatureMin}
 k_BT_{\rm min}(\omega)\approx 2\sqrt{2}\frac{k_{\rm B}T_{\rm j}L_{\rm sq}\omega}{R_j} \,.
\end{equation}

To our knowledge, this is the first time that the minimum noise temperature of a TC-optimized \dcsq{} has been calculated for an \emph{arbitrary} complex input impedance. However, Clarke \cite{clarke1979optimization} has previously determined the minimum noise-temperature of a \dcsq{} coupled to an input LCR resonator without a transformer. Clarke optimized the minimum \dcsq{} noise temperature by varying $C$ and $R$ of the resonator while holding the frequency and $L$ fixed, finding \cite{clarke1979optimization}
\begin{equation}\label{eq:ClarkeNoiseT}
k_{\rm B} T_{\rm min}=2.8 \frac{k_{\rm B}T_{\rm j}L_{\rm sq}\omega}{R_j}
\,,
\end{equation}
which is in close agreement with  Eq.~\ref{eq:TCNoiseTemperatureMin}. From Clarke's result combined with the calculation above, we conclude that a \dcsq{} can reach its global optimal minimum noise temperature when coupled to an LCR resonator of the optimal $Q$ and detuned resonance frequency.

We define $\eta$ as the ratio of the minimum noise temperature of an amplifier (with optimal complex impedance coupled to its input) to the quantum limit:
\begin{equation}\label{eq:eta}
    \eta(\omega)=\frac{k_BT_{\rm min}(\omega)}{\hbar\omega/2} \,.
\end{equation}
For a TC-optimized single-stage \dcsq{}, then, 
\begin{equation}\label{eq:TCeta}
 \eta_{\rm TC} \approx 4\sqrt{2}\frac{k_{\rm B}T_{\rm j}L_{\rm sq}}{\hbar R_j} \,,
\end{equation}
which is independent of the frequency $\omega$.

Eq.~\ref{eq:OnResonanceNoiseRatio} represents the ratio of the noise temperature of a TC-optimized \dcsq{} on resonance to the SQL, whereas Eq.~\ref{eq:TCeta} is the ratio to the SQL only for coupling to the optimal complex impedance. We will see in Section~\ref{sec:ScanSensitivityMatch} that $\eta$ is a useful parameter for scan sensitivity, including off-resonance operation.

\section{Amplifier chain}
\label{sec:squidchips}

Up to this point, we have presented an analysis of the noise of a single, weakly coupled \dcsq{} optimized according to Tesche and Clarke. In this section, we describe an implementation of a two-stage \dcsq{} chain and room-temperature preamplifier. We cite previous experimental results for the first-stage \dcsq{} and room-temperature preamplifier, and present here data taken with a prototype second-stage \dcsq{} series array designed to inform this analysis. In Section~\ref{sec:combo}, we analyze the contribution of all stages to the noise and the bandwidth.

The \dcsq{} amplifier chain consists of a tunable transformer to optimize the noise match of the \dcsq{} to the input circuit, a voltage-biased first-stage single \dcsq{} optimized according to TC theory, and a second-stage \dcsq{} array with $50\Omega$ output impedance that couples to a room-temperature preamplifier with $50\Omega$ input impedance (see Fig.~\ref{fig:ResonantCoupling}), allowing high bandwidth operation over a limited dynamic range.

The design drivers  for such an amplifier chain (originally based on \DMRm requirements, which are described in Section~\ref{sec:ScanSensitivityMatch}) are the following:
\begin{enumerate*}[label=\alph*)]
    \item The dynamic range of the input signal is very small.
    \item The lowest signal frequency is above the 1/f knee of the SQUID amplifiers and preamplifier electronics.
	\item The input signal has a narrow bandwidth (set by the resonator on the input).
    \item Noise matching to the resonator on the input requires a weakly coupled input.
\end{enumerate*}

These drivers have important consequences in the design of the amplifier chain:
\begin{enumerate*}[label=(\roman*)]
    \item The small dynamic range of the input signal permits open-loop operation of the SQUID amplifier chain.  We can use a simple, slow flux feedback to stabilize its gain and impedance match between amplifier stages.
	\item The small dynamic range also implies that the second-stage SQUID can be maintained at optimal ($50\Omega$) impedance match to the room-temperature preamplifier through its full signal swing, helping to mitigate signal reflections and optimize bandwidth.
    \item The relatively high lowest signal frequency is above the $1/f$ noise of the \dcsq{}, and avoids backgrounds that could prevent low dynamic range operations.
	\item The narrow bandwidth signal can help mitigate external interferences affecting the SQUID operation.
	\item The weak input coupling allows the use of an input coil with low intracoil capacitance and low capacitive coupling to the first-stage SQUID, enabling simple first-stage SQUID performance consistent with TC theory, and limiting the degradation of $Q$ from coupling to the loss in the \dcsq{}.  
 \end{enumerate*}

Most modern \dcsq{} applications require efficient coupling between the input circuit and the SQUID loop in order to maximize signal-to-noise ratio (SNR). 
In many of these more efficiently coupled \dcsq{}s, the increased coupling leads to interactions between the input circuit and the SQUID itself both at signal frequencies and at Josephson frequencies \cite{tesche1982optimization,tesche1983analysis,martinis1985signal,carroll1991small}. When input coils are efficiently coupled, non-ideal effects including intracoil capacitance, capacitance between the input coils and the SQUID, internal resonances, and inductive screening of the \dcsq{} lead to deviations from predicted TC noise performance. 
The implications of the imprecision, backaction, and correlation terms have been previously considered when the SQUIDs are tightly coupled to the resonator \cite{clarke1979optimization, koch1981quantum, tesche1982optimization, tesche1983analysis,martinis1985signal,carroll1991small}. They have also been considered in the weakly coupled limit, showing some of the same scaling that we find, but with SQUIDs that are not optimized according to TC theory \cite{levitin2007nuclear}.

To date, the full implications of these imprecision and backaction noise terms and their correlations over the frequency range up to 300\;MHz has not been referred to noise performance in a weakly coupled high-$Q$ resonator. This analysis is presented here.

The first- and second-stage \dcsq{}s as well as the room temperature preamplifier are based on designs that have been realized in practice.

\subsection{Tunable transformer}
\label{sec:TunableTransformer}

As seen in the previous section, the noise performance of the amplifier depends on the backaction and imprecision noise (and their correlations), which can be traded by tuning the coupling, $\kappa$, of the first stage SQUID to the high-\emph{Q} resonator.  In applications where the resonator frequency will be tuned, such as DMRadio-m$^3$, the optimal balance of imprecision and backaction noise may be different across different frequencies, and thus $\kappa$ is tunable \emph{in situ}.  We achieve this by implementing a SQUID input transformer consisting of a mechanically tuned variable mutual inductance.  We have demonstrated initial prototypes consisting of gradiometrically wound solenoidal coils on concentric sliding tubes, forming a variable mutual inductance, and actuated with a piezoelectric motors (similar to e.g. \cite{Boutan2018}).

\subsection{First-stage SQUID}
\label{sec:SQUID1description}
   
   The first-stage \dcsq{} considered here is based on SQUID C1 of \cite{wellstood1987low}, with additional details provided in \cite{wellstood1988excess}. This \dcsq{} has $I_0=6.3\,\mu$A, $R_{\rm j}=6\,\Omega$, and $L_{sq}=200$\,pH, and is weakly coupled to its input coil. In that work, it was shown that the TC theory is consistent with the measured white noise of the \dcsq{} when it is cooled to very low temperatures and the amplifier noise and low-frequency noise are subtracted. The resistive shunts in the \dcsq{} self-heat to $T_1 \approx 150$\,mK. Using Eq.~\ref{eq:OnResonanceNoiseRatio} with these parameters, it is seen that SQUID C1 of \cite{wellstood1987low} performs at approximately 9 times the SQL for optimal noise matching to a real impedance (e.g. on resonance), with no additional noise sources. From Eq.~\ref{eq:TCeta}, it is seen that $\eta_{\rm TC} \approx 3.7$, so SQUID C1 would perform at about 4 times the SQL for optimal noise matching to a complex impedance.

The construction of a C1-like \dcsq{} can be updated for a modern fabrication process. Updates include using a niobium/aluminum trilayer for the Josephson junctions in place of the Nb-NbO$_{\rm x}$-PbIn junction process originally used (described in \cite{wellstood1988excess}).

SQUID C1 was demonstrated up to 200\,MHz, and subsequent, similar SQUIDs were operated up to 300\,MHz \cite{Hilbert1985a}.  This paper considers frequency response to 300\,MHz (the top of the VHF band).

\subsection{Prototype second-stage SQUID} \label{sec:SQUID2description}

The second-stage \dcsq{} deviates significantly from an optimal TC \dcsq{}, because it is designed to operate reliably after the first stage of amplification, with parameters chosen to tolerate parasitic resonances with the input coil. Some internal microwave filters are implemented to damp these resonances and to block coupling between the SQUIDs at the Josephson frequency. The second-stage \dcsq{} is based on a series array of \dcsq{}s very similar to those used in the NIST series-array \dcsq{}s that have been deployed broadly in CMB, millimeter wave, and x-ray experiments worldwide \cite{ahmed2014bicep3, doriese2016developments,telescope2002scuba}. These \dcsq{}s are asymmetric and wound as serial gradiometers, minimizing coupling to environmental signals. Each input coil is a single turn, which reduces the parasitic capacitance.

 We conducted detailed measurements of the performance of a prototype second-stage \dcsq{} array, consisting of a series configuration of 20 \dcsq{}s, and extracted the necessary parameters for our analysis. These measurements form the basis for our design of the second-stage \dcsq{} arrays.
This SQUID has an asymmetric design, which provides internal self-feedback by coupling the output SQUID current from the second-stage back into the second-stage SQUID loop. This self-feedback is negative on the shallow slope of the SQUID response (column ``- feedback slope''), and positive on the steep slope (column ``+ feedback slope''). In the SQUID chain under consideration, we use the steep, positive feedback slope since it optimizes both the responsivity and the dynamic resistance on the output for high-bandwidth performance.

The prototype was cooled in a liquid-helium bath to $T_2=4$\,K. The output of the SQUID was connected through low-resistance Cu twisted triple (including ground) to a commmercial Magnicon XXF-1 preamplifier. The output was further amplified and then digitized for analysis. All data was taken open-loop (some data was validated by comparison to flux-locked-loop measurements). The following parameters were measured and are tabulated in Table~\ref{tab:SQUID20}: $I_c$\,min is the minimum critical current of the two SQUID junctions in parallel (at $\Phi_0/2$ flux bias). $I_c$\,max is the maximum critical current of the two junctions in parallel (at zero flux bias). $M_{\rm FB\_{2}}$ is the mutual inductance of the feedback coil to each SQUID in this 2nd-stage SQUID array. $M_{\rm IN\_{2}}$ is the mutual inductance to the input coil of each SQUID. Both $M_{\rm IN\_2}$ and $M_{\rm FB\_2}$ refer to the mutual inductances of individual SQUIDs in the array, and are thus independent of the number of SQUIDs in the array. 

The following data was taken at an optimal bias point ($\Phi\approx\Phi_0/2$). P is the measured power dissipated in the full SQUID array. Preamp $V_{\rm n\_PA}$ is the voltage noise of the Magnicon room-temperature preamplifier used in the measurement. Preamp $I_{\rm n\_PA}$ is the current noise of the Magnicon room-temperature preamplifier used in the measurement. $\left.\partial \Phi / \partial I \right|_{20\times1}$, $\left.\partial V / \partial I\right|_{20\times1}$, and $\left. \partial V / \partial \Phi \right|_{20\times1}$ are the three partial derivatives of the output voltage $V$, output current $I$, and input flux $\Phi$ of the $20\times1$ array, with the unlabeled parameter held fixed.

The noise at the optimal bias is white above 1\,kHz with no obvious pickup lines. To accurately measure this level, the noise is averaged between 1\,kHz and 100\,kHz with an SR760 100\,kHz spectrum analyzer, and referred to a flux noise in the input of the second-stage SQUID. $\Phi_{n2}$ total is the total measured effective flux noise referred to the input of the second-stage SQUID array at $T_2=4$\,K. Preamp $\Phi_n$ is the Magnicon preamplifier input voltage- and current-noise (added in quadrature) referred to a flux noise at the input of the 2nd-stage SQUID at the optimal bias point. SQUID $\Phi_{\rm n\_{20\times1}}$ is the 2nd-stage SQUID flux noise with the preamplifier noise subtracted in quadrature.

\input{2ndStageTable.tex}

\subsection{Three second-stage SQUID versions} \label{sec:SQUID2versions}

Three different 2nd-stage \dcsq{} designs are considered in this work. Each is scaled from the measured parameters in Table~\ref{tab:SQUID20}. All designs are based on the positive feedback slope (the right column). For our scaled design, in addition to using the measured data, we take the input coil inductance of each individual second-stage component SQUID of $L_{\rm SQ2}\approx120$\,pH (calculated using WRSpice \cite{wrspice}), and the estimated inductance from the wirebonds and interconnect to the first-stage SQUID, $L_{\rm int}$, which depends on the details of the packaging. Here, we take $L_{\rm int}\approx 2$\,nH.

The three designs will have different numbers of SQUIDs connected in series ($N_{\rm ser}$) and in parallel ($N_{\rm par}$). The total number of \dcsq{}s in the array is thus $N_{\rm SQ} = N_{\rm ser}\times N_{\rm par}$. The standard NIST SQUID array has 384 SQUIDs wired in different series and parallel configurations. Experience has shown that, when resonances are avoided, the dynamic parameters of these SQUIDs scale well in configurations combining both series and parallel arrays, up to 384 SQUIDs (more than the largest array presented here, which has 144). We have found that the parameters in Table~\ref{tab:SQUID20} scale as theoretically expected:

\begin{equation}\label{eq:PScaling}
    P \propto N_{\rm ser}\times N_{\rm par} 
\end{equation}
\begin{equation}
    \left.\frac{\partial \Phi}{\partial I}\right|_V   \propto \frac{1}{N_{\rm par}}
\end{equation}
\begin{equation}\label{eq:RdynScaling}
    \left.\frac{\partial V}{\partial I}\right|_\Phi   \propto {\frac{N_{\rm ser}}{N_{\rm par}}}
\end{equation}
\begin{equation}
    \left.\frac{\partial V}{\partial \Phi}\right|_I   \propto {N_{\rm ser}}
\end{equation}
\begin{equation}\label{eq:NoiseScaling}
\Phi_{\rm n_{SQUID}}^2 \propto \frac{T_2}{N_{\rm ser} N_{\rm par}}.
\end{equation}
Here, $\Phi_{\rm n_{SQUID}}$ refers to the equivalent imprecision flux noise in the SQUID array itself with an open input, and without any contributions from follow-on amplification. Eq.~\ref{eq:NoiseScaling} also shows scaling with $T_2$, the temperature of the second-stage \dcsq{} array. We find that this linear scaling is maintained to at least as low as 1\,K. At well below 1\,K, self-heating of the electrons in the \dcsq{} shunt resistor can limit further cooling.

All three 2nd-stage \dcsq{} array options are designed to have 50\,$\Omega$ dynamic output resistance at the optimal bias point. The arrays are scaled from the experimentally measured dynamic resistance of the prototype array that was tested. If the dynamic range of expected input signals is $\Delta \Phi \ll \Phi_0$, then the dynamic resistance should remain in this range. The output of the \dcsq{} is carried through a coaxial cable with 50\,$\Omega$ characteristic impedance to a room-temperature SQUID preamplifier with a low-noise, active 50\,$\Omega$ input impedance. No fast flux feedback is provided from room-temperature to the SQUIDs--they are operated in open-loop mode. It is expected that the bandwidth will be limited either by the coupling between 1st- and 2nd-stage SQUIDs, or by the bandwidth of the preamplifier circuit itself.

Using Table~\ref{tab:SQUID20} (biased on the positive feedback slope) and the scaling in Eq.~\ref{eq:RdynScaling}, the dynamic output resistance of a $16\times1$ series array of this design is $(16/20) \times 64.9$\,$\Omega \approx 50 $\,$\Omega$, which is the optimal dynamic resistance for high-bandwidth coupling to the output circuit, which consists of a coaxial cable with 50\,$\Omega$ characteristic impedance, and a room-temperature preamplifier with 50\,$\Omega$ active input impedance. The flux bias of the second-stage SQUID is periodically adjusted to maintain the dynamic output resistance at $\approx50 $\,$\Omega$ to maximize the bandwidth of the coupling to the preamplifier and to minimize signal reflections. Expected signals are small and should not appreciably change the dynamic output resistance. Second-stage \dcsq{} arrays of dimension $32\times2$ and $48\times3$ will also have $\approx50 $\,$\Omega$ output dynamic resistance, so that all three of these designs are well matched to the same output circuit. 

An example set of the three different second-stage SQUID designs to cover a range of bandwidths with minimal noise are determined from Table~\ref{tab:SQUID20} using the scaling in Eqs. \ref{eq:PScaling}--\ref{eq:NoiseScaling}. 
The three second-stage SQUID arrays considered here are:\\  

\begin{enumerate}
    \item High Bandwidth: a $16\times1$ series array SQUID. This series-array SQUID is identical to the measured SQUID in Table~\ref{tab:SQUID20} except that it has $N_{\rm ser}=16$ active SQUIDs in series instead of 20. All of the input coils are wound in series. The total input self-inductance of this design is $L_{2}=N_{\rm ser}N_{\rm par} L_{\rm SQ2} +L_{\rm dummy} +L_{\rm int}$.  Here $L_{\rm dummy} = 2N_{\rm par} L_{\rm SQ2}$ represents the additional total inductance from the input coils of one `dummy' SQUID at each end of every parallel bank of SQUIDs.  These `dummies' are SQUIDs made with open junctions in order to symmetrize the array; they are not connected on the output and thus do not contribute power dissipation or noise.  The resulting total input inductance is $L_2=18 L_{\rm SQ2} +L_{\rm int} \approx 3$\,nH.
    \item Medium Bandwidth: a $32\times2$ series array SQUID. This second-stage consists of $N_{\rm par}=2$ banks of $N_{\rm ser}=32$ series array SQUIDs, with the outputs of the two banks wired in parallel. The inputs of all SQUIDs are wound in series. The total input self-inductance of this design is $L_{\rm 2}=68 L_{\rm SQ2} +L_{\rm int}$. (68 instead of 64 because of the two dummy SQUIDs at the end of each series array).
    \item Low Bandwidth: a $48\times3$ series array SQUID. This second-stage consists of $N_{\rm par}=3$ banks of $N_{\rm ser}=48$ series array SQUIDs, with the outputs of the two banks wired in parallel. The inputs of all SQUIDs are wound in series. The total input self-inductance of this design is $L_{2}=150 L_{\rm SQ2} +L_{\rm int}$.
\end{enumerate}

\begin{table}[]
\centering
\begin{tabular}{c|ccc|}
\multicolumn{1}{l|}{} & Low BW                                 & Medium BW                             & High BW          \\ \hline
\# SQUIDs               & \multicolumn{1}{c|}{144 ($48\times3$)} & \multicolumn{1}{c|}{64 ($32\times2$)} & 16 ($16\times1$) \\
$R_{\rm dyn}$         & \multicolumn{1}{c|}{50 $\Omega$}       & \multicolumn{1}{c|}{50 $\Omega$}      & 50 $\Omega$      \\
P                    & \multicolumn{1}{c|}{6.00\,nW}           & \multicolumn{1}{c|}{2.67\,nW}         & 0.67\,nW         \\
$L_{2}$                & \multicolumn{1}{c|}{20\,nH }                 & \multicolumn{1}{c|}{10\,nH}               & 4\,nH               \\
$\tau=L_{2}/R_{\rm dyn 1}$       & \multicolumn{1}{c|}{3.3\,ns}         & \multicolumn{1}{c|}{1.7\,ns}       & 0.7\,ns       
\end{tabular}
\caption{Three different two-stage SQUID designs considered in this work (low, medium, and high bandwidth). The table includes the number of SQUIDs unit cells in the array $N_{\rm ser}\times N_{\rm par}$, dynamic resistance $R_{\rm dyn}$, power dissipation P, input-coil self inductance $L_2$, and time constant $\tau$ for coupling between first- and second-stage SQUIDs, using $R_{\rm dyn1}=6$\,$\Omega$.  All three designs use the same SQUID unit cell as the prototype $20\times1$ second-stage SQUID with measured parameters in Tbl. \ref{tab:SQUID20}. And all SQUID parameters are scaled according to Eqs.~\ref{eq:PScaling}--\ref{eq:NoiseScaling}. }\label{tab:2ndSQUIDparameters}
\end{table}

In Table~\ref{tab:2ndSQUIDparameters}, we present the dynamic resistance, power dissipation, self-inductance $L_2$, and also the time constant limiting the bandwidth of the connection between the first- and second-stage SQUIDs, $\tau=L_2/R_{\rm dyn1}$, where $R_{\rm dyn1}=6$\,$\Omega$ is the output dynamic resistance of the first-stage SQUID described in Section~\ref{sec:SQUID1description}.

\subsection{Room-temperature Op-amp Mode SQUID preamplifier}
\label{sec:RT_preamp}
At frequencies below 50\,MHz, a commercial ``op-amp mode'' preamplifier, the Magnicon XXF-1 \cite{MagniconXXF1}, is suitable for reading out the second-stage SQUID. These preamplifier electronics have open loop bandwidth of 50\,MHz, voltage noise of 0.33\,nV\,Hz$^{1/2}$, input current noise of 2.6\,pA\,Hz$^{1/2}$, and low-noise input impedance of 50\,$\Omega$, established by internal negative feedback from the output of the first transistor stage. 

At frequencies above 50\,MHz, we consider an ``op-amp mode'' preamplifier based on the ultra-high-speed SQUID electronics described in \cite{drung2003high,drung2005dc}, which was designed for small-signal open-loop bandwidth of 300\,MHz. These electronics utilize discrete bipolar rf transistors to achieve low-noise operation up to 300\,MHz. It achieves voltage noise of 0.3\,nV\,Hz$^{1/2}$, input current noise of 6\,pA\,Hz$^{1/2}$, and low-noise input impedance of 50\,$\Omega$, established by internal negative feedback from the output of the first transistor stage. 

\subsection{Cryogenic RF amplifier}
\label{sec:RF_preamp}

We also consider an alternative implementation in which a commercial cryogenic ``scattering mode'' RF amplifier is used for high-frequency measurements, and is diplexed to a Magnicon XXF-1 for low-frequency SQUID characterization and biasing.

One example of such an RF amplifier is the \CMT{} produced by Cosmic Microwave Technology, Inc \cite{CMTref}. It possesses an input impedance of 50\,$\Omega$, so it is impedance matched to the 50\,$\Omega$ output impedance of all second-stage SQUID designs considered here. The \CMT{}  has a gain of 32 dB over a 5--500\;MHz frequency range. It achieves a noise temperature of less than $T_{\rm n\_RF}=2$\,K from 5 to 500\;MHz. To measure the low-frequency response of the second-stage SQUID array, and to maintain the bias point with low-frequency flux feedback, the RF amplifier can be preceded by a diplexer that routes the information at and near DC to Magnicon XXF-1 electronics. 

\section{SQUID chain noise and bandwidth}
\label{sec:combo}
As discussed in Section~\ref{sec:squidnoise} and Section~\ref{sec:NoiseMatching}, the noise performance of the first-stage SQUID is determined by its imprecision noise current, backaction noise voltage, and their correlations.  The backaction noise voltage is dominated by the contribution of the first-stage SQUID, but the imprecision noise includes contributions from the full \dcsq{} chain -- not only the first-stage \dcsq{}, but also the second-stage \dcsq{}, the room-temperature preamplifier, and Johnson noise from normal resistance in the wires between the second-stage \dcsq{} and the preamplifier. The noise contributions from the later stages, which are uncorrelated with the first-stage SQUID noise, are referred to the input of the first stage where they add in quadrature to the imprecision current noise $S_{II}$ of the first-stage \dcsq{} (as given by Eq.~\ref{eq:ImprecisionCurrent}). 

Although the second-stage \dcsq{} and preamp stages backact on the first-stage \dcsq{}, their backaction does not significantly increase the physical fluctuation currents in the input \dcsq{} loop at the impedances and frequencies under consideration. This is because the output resistance of the first-stage \dcsq{} is much higher than the noise impedance of the second-stage SQUID. Therefore, we will assume the later stages do not modify the first-stage backaction noise $S_{VV}$ (Eq.~\ref{eq:BackactionVoltage}). Since the referred contributions of the later stages to the imprecision noise of the first stage are uncorrelated with the first-stage backaction, they also do not significantly affect $S_{IV}$ (Eq.~\ref{eq:InputCorrelation}).

We begin by examining the added imprecision noise from the second-stage SQUID, and subsequently include the added imprecision noise from the leads to the room-temperature preamplifier and from the room-temperature preamplifier itself.

\subsection{Second-stage SQUID bandwidth and noise}

\noindent
\\\emph{Response bandwidth:}
The output of the second-stage SQUID is 50\,$\Omega$, and it is impedance matched to the characteristic impedance of the coaxial transmission line and the input impedance of the room-temperature preamplifier, which is fast enough that the second-stage-to-preamp bandwidth is greater than the 300\,MHz target for the full system. The bandwidth of the first-stage input coil can be limited by parasitic capacitance, but it is weakly coupled in this implementation ($\kappa \ll 1$) so that it can also achieve the full-system 300\,MHz bandwidth. 

The bandwidth of the full SQUID chain is limited by the 1st-stage to 2nd-stage coupling, which is not impedance matched. The first-stage-to-second-stage SQUID coupling introduces a low-pass single-pole rolloff with time constant $\tau=L_{2}/R_{\rm dyn}$, 
 The time constants of the three SQUID chains considered here are listed in Table~\ref{tab:2ndSQUIDparameters}, ranging from 0.5\,ns to 3\,ns.\\

This time constant can be used to refer the second-stage SQUID flux noise $\Phi_{n2}$ to a first stage input imprecision current noise as a function of frequency $\omega$:
\begin{equation}\label{eq:2ndIto1stI}
\begin{aligned}
    \mathcal{S}_{II\_{\rm ref}(\omega)}&= \frac{(\Phi_{\rm n2})^2}{(M_{\rm IN\_2})^2 } \left(1 + \omega^2 \tau^2 \right) \bigg/ \left(M \left. \frac{\partial I_{\rm out}}{\partial \Phi} \right|_{V_{\rm out}} \right)^2  \,\\
    &= \frac{(\Phi_{\rm n2})^2}{(M_{\rm IN\_2})^2}\frac{ L_{\rm sq}}{ \kappa^2 L_{\rm in}} \left(1 + \omega^2 \tau^2 \right)  \,.
\end{aligned}
\end{equation}
Here we have used Eq.~\ref{eq:inputCoupling} for $M$ and Eq.~\ref{eq:FluxToCurrent} for the current responsivity of the first-stage \dcsq{}.

The noise of the second-stage SQUID itself and the noise of the follow-on preamplifier can be referred to an equivalent flux noise $\Phi_{\rm n2}$ in the second-stage SQUID. The noise components add in quadrature:

\begin{equation}\label{eq:Phisum}
    \Phi_{\rm n2}^2\approx \Phi_{\rm n\_20\times1}^2 \frac{20}{N_{\rm ser}N_{\rm par}}\frac{T_2}{4 \rm{K}}+\Phi_{\rm n2\_AMP}^2\,.
\end{equation}
The first term on the RHS is the flux noise of the second-stage SQUID itself, using Eq.~\ref{eq:NoiseScaling} to scale from the 20-dc SQUID array measurements in Table~\ref{tab:SQUID20} measured at 4\,K.
We expect to operate the second-stage SQUID at temperature $T_2=1$\,K, which has proven to be a stable operation temperature for \dcsq{}s of this design (see, e.g. \cite{Henderson2016}). 

The noise of the follow-on amplifier is referred to a flux noise in the second-stage SQUID, $\Phi_{\rm n2\_AMP}$ in Eq.~\ref{eq:Phisum}. This follow-on amplifier can be either a room-temperature op-amp mode SQUID preamplifier (Section~\ref{sec:RT_preamp_noise}), or a cryogenic RF amplifier (Section~\ref{sec:Cryo_RF_noise}).

\subsection{Noise of Op-amp Mode SQUID preamplifier}
\label{sec:RT_preamp_noise}

The referred second-stage flux noise of the op-amp mode room-temperature SQUID preamplifier has contributions from the noise of the room-temperature preamplifier plus Johnson noise of the leads to room temperature $\Phi_{\rm n2\_VPAL}^2$ (Eq.~\ref{eq:PreampV2}), and the current noise of the room-temperature preamplifier $\Phi_{\rm n2\_IPA}^2$ (Eq.~\ref{eq:PreampI2}), which are presumed to be uncorrelated:

\begin{equation}\label{eq:TotalOpampPA}
    \Phi_{\rm n2\_AMP}^2=\Phi_{\rm n2\_VPAL}^2 + \Phi_{\rm n2\_IPA}^2\,.
\end{equation} 

The voltage noise of the room-temperature preamplifier and the Johnson voltage noise of the leads to room temperature are referred to a flux noise in the second-stage SQUID array:
\begin{equation}\label{eq:PreampV2}
    \Phi_{\rm n2\_VPAL}^2\approx \frac{(V_{\rm n\_PA})^2 + 4 k_{\rm B}T_{\rm lead}R_{\rm lead}(\omega)}{ (\left.\partial V / \partial \Phi\right|_{20\times1}) ^2} \left(\frac{20}{N_{\rm ser}}\right)^2 \,.  
\end{equation}
Here, $V_{\rm n\_PA}$ is the preamp voltage noise. The voltage transfer function is increased proportional to the number of \dcsq{}s in series, $N_{\rm ser}$, according to Eq.~\ref{eq:RdynScaling}. $R_{\rm lead}$ and $T_{\rm lead}$ are the effective real resistance and temperature of the leads from the second-stage SQUID to the room temperature op-amp mode preamplifier. $R_{\rm lead}$ represents the cable loss at the operational frequency (which is generally higher than dc resistance). This noise is conservative, as the increase in the critical current of niobium films as they are cooled below 4\,K causes the flux-to-voltage transfer function to steepen slightly on cooling to 1\,K. For this analysis, we assume superconducting coaxial cable from the second-stage SQUID at 1\,K to where it is heat sunk at 4\,K. Copper coaxial cable from 4\,K to a room-temperature preamplifier, with $R_{\rm lead}=1$\,$\Omega$ (assumed frequency independent to simplify the analysis) and $T_{\rm lead}=200$\,K is assumed.

The current noise of the room-temperature preamplifier is referred to a flux noise in the second-stage SQUID array:
\begin{equation}\label{eq:PreampI2}
    \Phi_{\rm n2\_IPA}^2
    \approx  (I_{\rm n\_PA})^2
    \left. \left(\frac{  \partial \Phi }{\partial I}\right|_{20\times1} \frac{1}{N_{\rm par}}  \right)^2
     \,.
\end{equation}
Here, $I_{\rm n\_PA}$ is the preamp current noise and $ \left.\partial \Phi /\partial I\right|_{20\times1}$ is the current-to-flux transfer function for the 20-SQUID array. We assume that the resistance of the leads is small enough that a stiff current bias can be maintained from room temperature at HF and VHF signal frequencies.

\subsection{Noise of Cryogenic RF preamplifier}
\label{sec:Cryo_RF_noise}
We also consider an implementation with
a commercial cryogenic ``scattering mode” RF amplifier as an alternative to the ``op-amp mode" room-temperature preamplifier. Since the RF amplifier is operated at 4\,K, it is connected to the second-stage \dcsq{} through a superconducting coaxial cable, so we do not include any Johnson noise between the second-stage SQUID and the cryogenic RF amplifier. The noise temperature $T_{n\_RF} \approx 2$K of the cryogenic RF amplifier with 50\,$\Omega$ input impedance can be referred to an equivalent voltage noise at the output of the second-stage SQUID. We may refer this voltage noise to a flux noise in the second-stage SQUID by matching the dynamic resistance of the second stage, $\left.\partial{V}/\partial{I}\right|_V  = 50\,\Omega$ with a matched $50\,\Omega$ RF amplifier input impedance, and using Euler's chain rule (see App. \ref{sec:Euler}):
\begin{equation}\label{eq:rfampnoise}
\begin{split}
    \Phi_{\rm n2\_AMP}^2\approx - 4& k_{B}T_{\rm n\_RF} 
    \left.  \frac{\partial \Phi}{\partial I_{\rm out}} \right|_{20\times1} \\
    &\times \left. \frac{\partial \Phi}{\partial V_{\rm out}} \right|_{20\times1} 
    \left(\frac{20}{N_{\rm ser}N_{\rm par}}\right) \,. 
\end{split}
\end{equation}

By substituting Eq.~\ref{eq:rfampnoise} for a cryogenic follow-on amplifier, rather than Eq.~\ref{eq:TotalOpampPA}, into Eq.~\ref{eq:Phisum}, one can follow the same noise analysis as outlined above to determine the resulting total noise performance of the system. The cryogenic RF amplifier can similarly operate with the low-, medium-, or high-bandwidth versions of the second-stage SQUID (Table~\ref{tab:2ndSQUIDparameters}).

\subsection{Full system noise}
\label{sec:full}
We can now compute the full-system imprecision current noise referred to the input of the resonator circuit as
\begin{equation}\label{eq:fullimprecision}
    \mathcal{S}_{II_{\rm sys}}(\omega)= \mathcal{S}_{II}+\mathcal{S}_{II\_{\rm ref}(\omega)} \,,
\end{equation}
where $\mathcal{S}_{II}$ has the TC value of imprecision current noise from Eq.~\ref{eq:ImprecisionCurrent}, and $\mathcal{S}_{II\_{\rm ref}(\omega)}$ is the referred noise from the follow-on amplifier chain from Eq.~\ref{eq:2ndIto1stI} and Eq.~\ref{eq:Phisum}.

\begin{figure}[t]
    \centering
    \includegraphics[width=\linewidth]{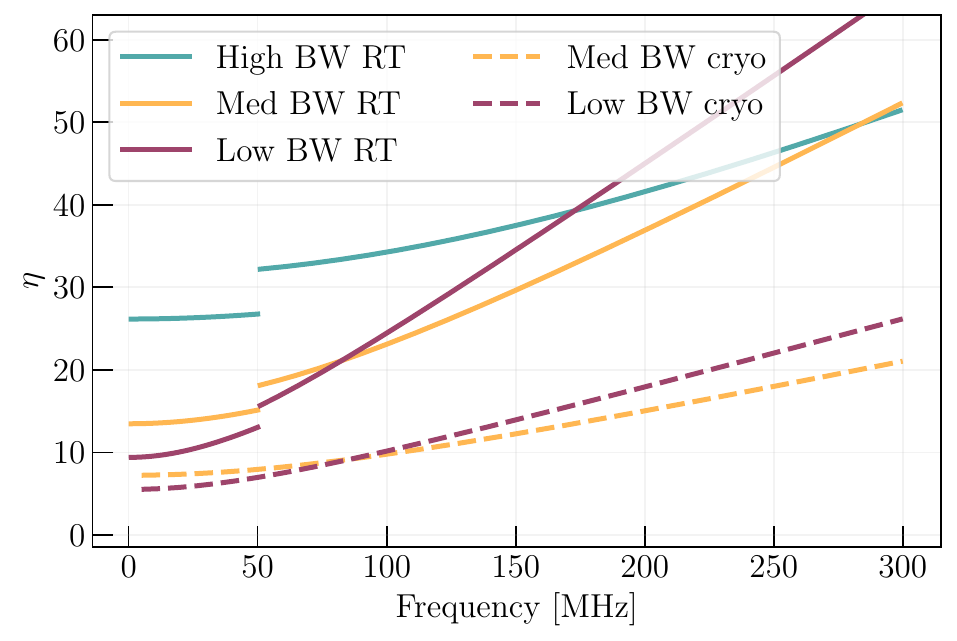}
    \caption{Noise parameter $\eta$ as a function of frequency for the full \dcsq{} amplifier chain (from Eq.~\ref{eq:FullNoiseTemperatureComplexMatch}). Blue lines are the high-bandwidth 2nd-stage SQUID, purple are the medium-bandwidth, and yellow are the low-bandwidth (see Table~\ref{tab:2ndSQUIDparameters}). Solid lines use room-temperature  (RT) op-amp mode amplifiers: Magnicon XXF-1 below 50\,MHz and the high bandwidth room temperature preamp above 50\,MHz (Section~\ref{sec:RT_preamp}). Dashed lines use the \CMT{} cryogenic RF preamp (Section~\ref{sec:RF_preamp}) above 10\,MHz. The dashed line with the high BW 2nd-stage SQUID with the cryogenic RF preamp is omitted for clarity; this 2nd-stage SQUID option might be useful only at frequencies above 300\,MHz.  As discussed in \cite{alshirawi2023electromagnetic}, the low- and medium-bandwidth configurations have sufficiently low $\eta$ to serve DMRadio-m$^3$ to reach its axion sensitivity goals.}
    \label{fig:EtaPlot1}
\end{figure}

As discussed in Section~\ref{sec:NoiseTempComplex}, the minimum noise temperature that can be achieved requires an optimal complex input impedance connected to the SQUID input (see Eq.~\ref{eq:TCNoiseTemperatureMin} for a computation of the minimum noise temperature for a TC-optimized first-stage SQUID). This noise temperature will not be achieved with a real input impedance, which can not take advantage of the fact that the backaction and imprecision noise are correlated, but it is still a useful parameter. The minimum full-system noise temperature of the full amplifier chain connected to an optimal complex input impedance is calculated from Eq.~\ref{eq:TCNoiseTemperatureComplexMatch}. The value of $\eta$ can then be computed as the ratio of this noise temperature to the SQL for added noise, $\hbar\omega/2$. Once again, we take $\operatorname{Re}\{\mathcal{S}_{IV}\}=0$, so that 
\begin{equation}\label{eq:FullNoiseTemperatureComplexMatch}
 \eta(\omega)=\frac{k_BT_{\rm min}(\omega)}{\hbar\omega/2}=\frac{1}{\hbar\omega}\sqrt{\mathcal{S}_{VV}\mathcal{S}_{II_{\rm sys}}(\omega) - (\operatorname{Im}\{\mathcal{S}_{IV}\})^2}\,,
\end{equation}
where the TC value of $\mathcal{S}_{VV}$ is used from Eq.~\ref{eq:BackactionVoltage}, the TC value of $\mathcal{S}_{IV}$ is used from Eq.~\ref{eq:InputCorrelation}, and the full-system imprecision current noise $\mathcal{S}_{II_{\rm sys}}(\omega)$ is determined by Eq.~\ref{eq:fullimprecision}. For our assumed SQUID chain parameters, we plot the value of $\eta(\omega)$ for the three second-stage SQUID versions up to 300\,MHz in Fig.~\ref{fig:EtaPlot1}. In the three solid lines, we use a Magnicon XXF-1 preamplifier below 50\,MHz. Above 50\,MHz, we assume the use of the room-temperature op-amp mode preamplifier described in Section~\ref{sec:RT_preamp}. Different second-stage SQUIDs are optimal from a noise perspective at different frequency ranges. 
The dashed lines were attained using the cryogenic RF preamplifier and the low- and medium-bandwidth second-stage SQUID variety, which seems to be optimal (or close to optimal) at all relevant frequencies for the cryogenic RF amplifier.

\begin{figure}[t]
    \centering
    \includegraphics[width=\linewidth]{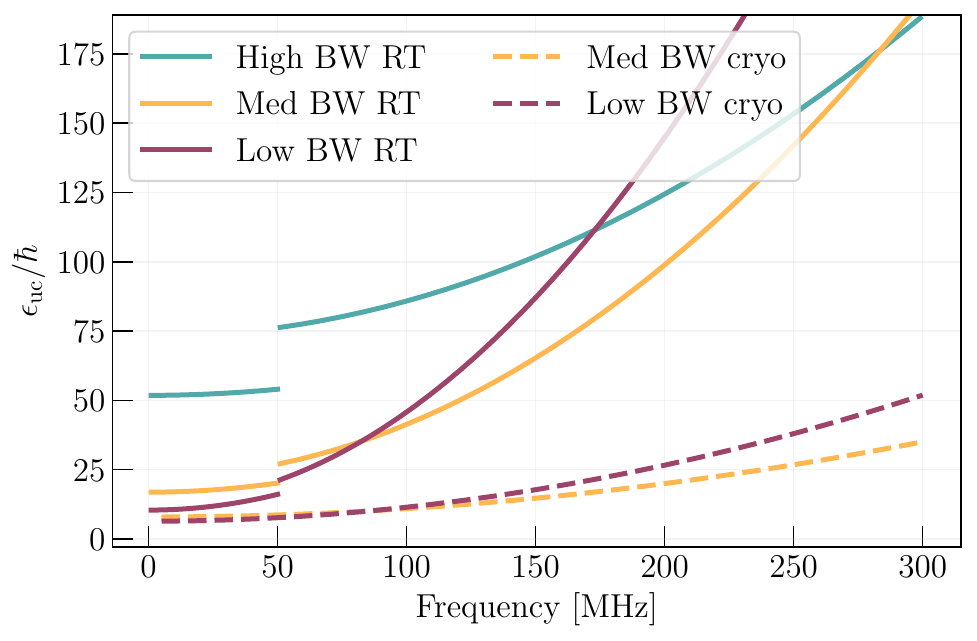}
    \caption{The uncoupled energy sensitivity $\epsilon_{\rm uc}$ (in units of $\hbar$) as a function of frequency for the full \dcsq{} amplifier chain (from Eq.~\ref{eq:FullSystemEpsilonImprecision}). Details of all lines are the same as in Fig.~\ref{fig:EtaPlot1} caption. $\epsilon_\text{uc}$ for the low-bandwidth SQUID with the cryogenic RF preamp has values less than $7.6\hbar$ below 50  MHz.}
    \label{fig:EpsilonPlot1}
\end{figure}

The full-system uncoupled imprecision energy sensitivity can be computed from the full system current noise
\begin{equation}\label{eq:FullSystemEpsilonImprecision}
    \epsilon_{\rm uc}(\omega)= \frac{M^2\mathcal{S}_{II_{\rm sys}}(\omega)}{2 L_{\rm sq}}=\frac{\kappa^2L_{\rm in}\mathcal{S}_{II_{\rm sys}}(\omega)}{2},
\end{equation}
where $\mathcal{S}_{II_{\rm sys}}$ is from Eq.~\ref{eq:fullimprecision}, which contains both the first-stage 
TC contribution calculated in Eq.~\ref{eq:EpsilonImprecision}, and the referred noise from the follow-on amplifiers. We plot $\epsilon_{\rm uc}$ for the three second-stage SQUID versions and the room-temperature amplifier, as well as the medium-bandwidth second stage with the cryogenic RF amplifier, in Fig.~\ref{fig:EpsilonPlot1}.

As will be seen in Section~\ref{sec:ScanSensitivityMatch}, the values of $\eta(\omega)$ and $\epsilon_{\rm uc}(\omega)$ computed in Eqs. \ref{eq:FullNoiseTemperatureComplexMatch} and \ref{eq:FullSystemEpsilonImprecision}, with results in Figs. \ref{fig:EtaPlot1} and \ref{fig:EpsilonPlot1}, determine the optimal matching conditions and the frequency-integrated scan sensitivity for resonant axion searches.

\section{SQUID Chain Noise Match on Resonance}
\label{sec:onres}

Eqs. \ref{eq:OnResonanceMatch} and \ref{eq:TCNoiseTemperature} describe the noise-match condition and the optimal on-resonance noise temperature for an ideal TC \dcsq{} with no follow-on amplifier noise. It is straightforward to extend this analysis to the full amplifier chain. On resonance, the input circuit presents a real impedance. Then, from Eq.~\ref{eq:OnResonanceNoiseImpedance}, the result is a resonance-frequency-dependent noise resistance
\begin{equation}\label{eq:FullOnResonanceNoiseImpedance2}
R(\omega_0)=\sqrt{\frac{ \mathcal{S}_{VV}}{ \mathcal{S}_{II}+\mathcal{S}_{II\_{\rm ref}(\omega_0)}}} \,,
\end{equation}
where the backaction $\mathcal{S}_{VV}$ arises from the TC-optimized first-stage SQUID backaction voltage (Eq.~\ref{eq:BackactionVoltage}), $\mathcal{S}_{II}$ is the TC-optimized first-stage SQUID imprecision noise current (Eq.~\ref{eq:ImprecisionCurrent})  and the follow-on amplifier referred noise
 $\mathcal{S}_{II\_{\rm ref}(\omega)}$ is from Eq.~\ref{eq:2ndIto1stI} and Eq.~\ref{eq:Phisum}. As was previously done for the TC first-stage SQUID in Section~\ref{sec:onresonance}, Eq.~\ref{eq:FullOnResonanceNoiseImpedance2} can be solved for the value of $\kappa_{\rm g}$ that is noise matched, achieving the minimum noise temperature. The value of $\kappa_{\rm g}$ can be tuned with a variable input transformer.

Similarly, from Eq.~\ref{eq:OnResonanceMinNoiseTemperature}, we use the full system imprecision noise to arrive at the minimum on-resonance noise-temperature that can be achieved with the full amplifier chain:
\begin{equation}\label{eq:OnResonanceMinNoiseTemperature2}
k_{\rm B}T_{\rm min}(\omega_0)=\frac{1}{2} \sqrt{ \mathcal{S}_{VV}(\mathcal{S}_{II}+\mathcal{S}_{II\_{\rm ref}(\omega_0)})}.
\end{equation}
These parameters can be computed using the derivations in Section~\ref{sec:combo}. 

It is evident from Eq.~\ref{eq:FullOnResonanceNoiseImpedance2} that the additional imprecision noise from the follow-on amplifiers decreases the noise impedance of the amplifier, which can be compensated for by tuning the transformer to a stronger coupling (higher $\kappa_{\rm g}$ and mutual inductance $M$). From Eq.~\ref{eq:OnResonanceMinNoiseTemperature2}, the addition of the follow-on amplifier noise 
increases the minimum noise temperature at this new optimal match.

\section{Scan Sensitivity Noise Match}
\label{sec:ScanSensitivityMatch}

In Section~\ref{sec:onres}, we described the matching condition for optimal noise temperature on resonance, where the resonator impedance is real. In on-resonance applications, when the coupling is optimized, the imprecision noise current is equal to the backaction-driven noise current. On resonance, imaginary correlations do not affect the total noise, which is a quadrature sum of the backaction and imprecision. The presence of a thermal population of photons in the resonator does not affect the on-resonance noise-match condition.

In some applications, however, information is used that is detuned from the resonance frequency. At detuned frequencies, the backaction-driven current is rolled off by the Lorentzian lineshape, while imprecision noise is not. The presence of a thermal photon population, which is also rolled off by the Lorentzian, can significantly change the optimal match condition. Furthermore, the resonance has a complex impedance, so imaginary \dcsq{} correlations have a significant effect on the noise, either increasing (on one side of the Lorentzian) or decreasing it (on the other side) \cite{Chaudhuri:2018rqn}. 

One example of such an application is axion searches.  The mass/frequency of the axion is unknown, so the primary figure of merit is not signal-to-noise ratio at resonance, but the integrated scan sensitivity, i.e. the square of SNR integrated over all frequencies and summed over all tuning steps. The calculation of the integrated scan sensitivity using an arbitrary flux-to-voltage amplifier with correlations is worked out in \cite{Chaudhuri:2018rqn} (Eq.~F58), and the full analysis of \DMRm{} sensitivity has been calculated using this formalism in
\cite{alshirawi2023electromagnetic}.  After a \dcsq{} amplifier chain is matched through a tunable transformer to an input resonator to optimize scan sensitivity, only the noise parameter $\eta(\omega)$ is needed to determine the SNR of the axion search. We leave the full scan sensitivity calculation to \cite{Chaudhuri:2018rqn} and here just state the conclusions. 

The optimal noise-match condition for scan sensitivity, defined by the total system imprecision noise, is (Eq.~F55 in \cite{Chaudhuri:2018rqn})
\begin{equation}\label{eq:SII_scanopt}
    \mathcal{S}_{II{\rm sys}}=
    \frac{\hbar\omega}{R} \frac{2\eta^{2}}{2n+1 + \sqrt{(2n+1)^{2} + 8\eta^{2}}},
\end{equation}
where $n$ is the resonator's thermal occupation number,
\begin{equation}
    n=\frac{1}{\exp(\hbar\omega/k_{\rm B}T_{\rm res})-1}.
\end{equation}
Using Eq.~\ref{eq:FullSystemEpsilonImprecision} to express $\mathcal{S}_{II{\rm sys}}$ in terms of $\epsilon_{\rm uc}$, and also using the definition of $\kappa_{\rm g}$ from Eq.~\ref{eq:kappaG}, we solve for the optimal matching condition on $\kappa_{\rm g}$:
\begin{equation}\label{eq:ScanMatching}
\kappa_{\rm g}^2=\frac{1}{Q}\frac{\epsilon_{\rm uc}}{\hbar } \frac{2n+1 + \sqrt{(2n+1)^{2} + 8\eta^{2}}}{\eta^{2}}.
\end{equation}

The matching condition in Eq.~\ref{eq:ScanMatching}  optimizes the scan sensitivity for a Lorentzian resonator, including the effect of the correlations. As the thermal occupation of the resonator $n$ increases in Eq.~\ref{eq:ScanMatching}, the coupling $\kappa_{\rm g}$ must also increase to optimize the scan sensitivity. We further see that $\kappa_{\rm g}^2 \propto 1/Q$, so that high-$Q$ resonators require low global coupling efficiency. In the case of \DMRm{}, the optimal coupling efficiency is low enough for the use of the weakly coupled SQUID design C1 of \cite{wellstood1987low}, whose noise was accurately described by TC theory.

The noise properties of a \dcsq{} amplifier chain for a scanning experiment such as an axion search are completely specified by the two parameters $\eta$ and $\epsilon_{\rm uc}$, which we have presented in this paper for both TC-optimized single SQUIDs (Eqs. \ref{eq:TCeta} and \ref{eq:EpsilonImprecision}), and the full, high-bandwidth \dcsq{} amplifier chain (Figs.  \ref{fig:EtaPlot1} and \ref{fig:EpsilonPlot1}).




 \section{Conclusions}
\label{sec:conclusion}

In this paper, we show the limits on noise for the readout of high-$Q$ resonators below 300\,MHz using dc SQUIDs optimized according to Tesche \& Clarke theory, including practical implementations with follow-on amplifier chains. These analyses take into account SQUID backaction noise, imprecision noise, and their correlations.
In Eq.~\ref{eq:TCNoiseTemperature}, we show the minimum noise temperature achievable with a TC-optimized first-stage SQUID on resonance.  We also find the minimum noise temperature that can be achieved with a matched complex impedance in Eq.~\ref{eq:TCNoiseTemperatureMin}, show that this noise temperature can be achieved with a detuned LCR resonator, and compare to the quantum limit in Eq.~\ref{eq:TCeta}. 

We present data from a realized prototype second-stage SQUID and show how it scales to three different second-stage SQUID designs with different bandwidth. We analyze the properties of several full amplifier chain designs constructed of realized components: a demonstrated TC-optimized first-stage SQUID, three second-stage SQUID designs scaled from the realized prototype presented in this paper, the commercial Magnicon XXF-1 room-temperature preamplifier and \CMT{} cryogenic
RF preamp, and the 300\,MHz room-temperature preamplifier described in Section~\ref{sec:RT_preamp}. We analyze bandwidth, backaction noise, imprecision noise, and correlations.

We calculate how to optimally match these SQUID amplifier chains to a resonator to maximize scan sensitivity for axion searches, and we provide the two parameters, $\eta$ and $\epsilon_{\rm uc}$, that specify the achieved noise and matching conditions. The values of these two parameters are shown as a function of frequency for several versions of the SQUID amplifier chain in Figs. \ref{fig:EtaPlot1} and \ref{fig:EpsilonPlot1}.

This analysis is of importance to any application that requires optimal matching of weakly coupled dc SQUIDs to low-temperature, high-$Q$ resonators below 300\,MHz. The design is the basis for the SQUID amplifier chain for the DMRadio-m$^3$ axion search, and is used for its full electromagnetic design and science reach analysis in \cite{alshirawi2023electromagnetic}.

\section{Acknowledgments}
\label{sec:Acknowledgments}

The authors acknowledge support for DMRadio-m$^3$ as
part of the DOE Dark Matter New Initiatives program
under SLAC FWP 100559. 
C. Bartram acknowledges support from
the Panofsky Fellowship at the SLAC National Accelerator Laboratory. S. Chaudhuri acknowledges support
from the R. H. Dicke Postdoctoral Fellowship and Dave
Wilkinson Fund at Princeton University. 
C. P. Salemi was supported by the Kavli Institute for Particle Astrophysics
and Cosmology Porat Fellowship. J. Clarke acknowledges support from the U.S. Department of Energy, Office of Science, National Quantum Information Science Research Centers.

Certain equipment, instruments, software, or materials are identified in this paper in order to specify the experimental procedure adequately. Such identification is not intended to imply recommendation or endorsement of any product or service by NIST, nor is it intended to imply that the materials or equipment identified are necessarily the best available for the purpose.

%% file: 2ndStageTable.tex
\begin{table}[] 
\centering
\begin{tabular}{c|c|c} 
\multicolumn{1}{l|}{\textit{$20\times1$ SQUID}}                                    & \begin{tabular}[c]{@{}c@{}}- feedback \\ slope\end{tabular} & \begin{tabular}[c]{@{}c@{}}+ feedback \\ slope\end{tabular} \\ \hline
$I_c$ min                                                                          & 3.055 $\mu$A                                                & same                                                        \\
$I_c$ max                                                                          & 9.485 $\mu$A                                                & same                                                        \\
$M_{\rm FB\_{2}}$                                                                       & 44\,pH                                                      & same                                                        \\
$M_{\rm IN\_2}$                                                                       & 105\,pH                                                     & same                                                        \\
P                                                                                  & 0.834\,nW                                                   & same                                                        \\
Preamp $V_{\rm n\_PA}$                                                                       & 0.32 $\frac{\rm nV}{\sqrt{\rm Hz}}$                         & same                                                        \\
Preamp $I_{\rm n\_PA}$                                                                       & 2.7 $\frac{\rm pA}{\sqrt{\rm Hz}}$                          & same                                                        \\
$\left.\partial \Phi / \partial I \right|_{20\times1}$                                                       & 231\,pH                                                     & 95.5\,pH                                                    \\
$\left.\partial V / \partial I\right|_{20\times1}$                                                          & 154 $\Omega$                                                & 64.9 $\Omega$                                               \\
$\left. \partial V / \partial \Phi \right|_{20\times1}$                                                       & \multicolumn{1}{l|}{$6.66\times10^{11}$\,Hz}                & \multicolumn{1}{l}{$6.80\times10^{11}$\,Hz}                 \\ \hline
\begin{tabular}[c]{@{}c@{}}$\Phi_{n2}$ total\\ at $T_2=4$\,K\end{tabular}             & 0.453 $\frac{\mu \Phi_0}{\sqrt{\rm Hz}}$                    & 0.394 $\frac{\mu \Phi_0}{\sqrt{\rm Hz}}$                    \\ \hline
\begin{tabular}[c]{@{}c@{}}Preamp $\Phi_n$\\ referred\end{tabular}                 & 0.380 $\frac{\mu \Phi_0}{\sqrt{\rm Hz}}$                    & 0.259 $\frac{\mu \Phi_0}{\sqrt{\rm Hz}}$                    \\ \hline
\begin{tabular}[c]{@{}c@{}}SQUID $\Phi_{\rm n\_{20\times1}}$ \\ at $T_2=4$\,K\end{tabular} & 0.247 $\frac{\mu \Phi_0}{\sqrt{\rm Hz}}$                    & 0.297 $\frac{\mu \Phi_0}{\sqrt{\rm Hz}}$                   
\end{tabular}
\caption{Measured parameters of prototype series-array SQUID second stage. The SQUID array measured here consists of 20 SQUIDs in series. The measurements are taken at T=4\,K, at the bias point with optimum voltage-to-flux response (near $\Phi=\Phi_0/2$).
Dynamic parameters of the three 2nd-stage SQUIDs designs can be scaled from the parameters measured here.  }
\label{tab:SQUID20}
\end{table}

%% file: appendices.tex
\section{Response Functions}\label{sec:ResponseFunctions}

From Fig.~12b in \cite{tesche1977dc} (all image processing done using \cite{Rohatgi2022}), at optimally biased flux ($\Phi=\Phi_0/4$) and current ($I=1.8I_0$) and temperature figure of merit (FOM) $\Gamma=0.050$,
\begin{equation}
    \frac{\partial V}{\partial\Phi}=1.6 \frac{I_0R_j}{\Phi_0} \,.
\end{equation}
Assuming $\beta_L$ as in Eq.~\ref{eqn:TCassumptions}, we can rewrite the flux-to-voltage transfer function as
\begin{equation}
    \frac{\partial V}{\partial\Phi}=0.8 \frac{R_j}{L_{sq}} \,. 
\end{equation}
It should be noted that the original Tesche and Clarke paper \cite{tesche1977dc} had an incorrect voltage transfer function that was updated in \cite{Bruines1982}. By reducing the temperature (to get lower values of $\Gamma$), the transfer function scales as shown in Fig.~1c of \cite{Bruines1982}.  At current bias of $1.8I_0$, from $\Gamma=0.050\to0.025$, the transfer function scales by a factor of 1.15, giving
\begin{equation} \label{eq:dVdPhiacc}
    \frac{\partial V}{\partial\Phi}=0.9 \frac{R_j}{L_{sq}} \,
\end{equation}
for the lower temperature.  Although temperatures lower than $\sim4.2$~K ($\Gamma\approx0.025$) are not included in \cite{tesche1977dc} or \cite{Bruines1982}, modern cryogenics and SQUID package design can achieve considerably cooler operating temperatures leading to a prefactor closer to 1 in Eq.~\ref{eq:dVdPhiacc}.  We take
\begin{equation}
    \frac{\partial V}{\partial\Phi}\approx\frac{R_j}{L_{sq}} \,
\end{equation}
throughout this work.

Using similar bias current and flux, we can extract the dynamic resistance from Fig.~12a in \cite{tesche1977dc},
\begin{equation}
    \frac{\partial V}{\partial I} \approx 0.97R_j \approx R_j \,.
\end{equation}
We round the coefficient to unity for our calculations in the body of the paper, which is also the value used by Drung \cite{drung2003high}.  Note that this indicates an average voltage output of $0.38I_0R_j$.

\section{Euler's chain rule for \dcsq{}s}
\label{sec:Euler}

If the condition of current-bias at the Josephson frequency and voltage-bias at HF and VHF frequencies is obtained, then the relationship between the flux $\Phi$, voltage $V_{\rm out}$, and current $I_{\rm out}$ at VHF and HF frequencies can be expressed as a surface in $\Phi$, V, I space: $f(\Phi,V_{\rm out},I_{\rm out})=0$. At the TC optimal bias point, all of the partial derivatives of this function are nonzero, so that on this surface locally each variable is given as a differentiable function of the remaining two. Then, the triple product rule (Euler's chain rule) can be invoked
\begin{equation}\label{eq:Euler}
    \left. \frac{\partial I_{\rm out}}{\partial \Phi} \right|_{V_{\rm out}} \left. \frac{\partial \Phi}{\partial V_{\rm out}} \right|_{I_{\rm out}}\left. \frac{\partial V_{\rm out}}{\partial I_{\rm out}} \right|_{\Phi}=-1.
  \end{equation}
Additionally, using the relationship
\begin{equation}\label{eq:InverseDerivative}
\left. \frac{\partial V_{\rm out}}{\partial \Phi} \right|_{\Phi} \left. \frac{\partial \Phi}{\partial V_{\rm out}} \right|_{\Phi} = 1,
  \end{equation}
and Eqs. \ref{eq:FluxToVoltage} and \ref{eq:DynamicResistance}, we arrive at
\begin{equation}\label{eq:FluxToCurrentAppendix}
    \left. \frac{\partial I_{\rm out}}{\partial \Phi} \right|_{V_{\rm out}} \approx -\frac{1}{L_{sq}}.
  \end{equation}

%

\section{Cryogenic RF Preamplifier}\label{sec:RFAmp}
As an alternative to the ultra-high-speed room-temperature preamp described in Section~\ref{sec:RT_preamp}, we consider the use of a cryogenic RF amplifier to read out signal frequencies above 50 MHz. One example of such an RF amplifier is the \CMT{} produced by Cosmic Microwave Technology, Inc. It possesses an input impedance of 50\,$\Omega$, so it is impedance matched to the 50\,$\Omega$ output impedance of the second-stage SQUID. The \CMT{} has a gain of 32\;dB with a gain flatness of $\pm1$\;dB, and a noise temperature of better than $T_{\rm n\_RF}=2$\,K from 5--500\;MHz. To current bias the second-stage SQUID array and to maintain the bias point with low-frequency flux feedback, the RF amplifier may be preceded by a diplexer that routes the information at and near DC to the Magnicon XXF-1 electronics. 

We now calculate the values of $\epsilon_{\rm uc}$ and $\eta$ for the full SQUID chain previously shown in Figs. \ref{fig:EtaPlot1} and \ref{fig:EpsilonPlot1}. Since the RF amplifier is operated at 4\,K, it is connected to the second-stage \dcsq{} through a superconducting coaxial cable, and we take $R_{\rm lead}=0$. We can then refer the temperature noise of the cryogenic RF amplifier $T_{\rm n\_RF}$ to a flux noise in the second-stage SQUID:
\begin{equation}\label{eq:rfampnoise}
    \Phi_{\rm n2\_RF}^2\approx 4 k_{\rm B}T_{\rm n\_RF} \left(\frac{20}{N_{\rm ser}N_{\rm par}}\right)\left. \frac{\partial \Phi}{\partial I_{\rm out}} \right|_{20\times1}\left. \frac{\partial \Phi}{\partial V_{\rm out}} \right|_{20\times1} \,.  
\end{equation}
Then, the RF amp and second-stage SQUID noise referred to a flux in the input in the second-stage SQUID becomes:
\begin{equation}\label{eq:PhisumRF}
    \Phi_{\rm n2}^2=\Phi_{\rm n2\_SQ}^2+\Phi_{\rm n2\_RF}^2,
\end{equation}
where $\Phi_{\rm n2\_SQ}$ is still from Eq.~\ref{eq:Phisum}. Eq.~\ref{eq:2ndIto1stI} is then used to compute $\mathcal{S}_{II\_{\rm ref}(\omega)}$, and the rest of the analysis proceeds as already described in Section~\ref{sec:squidchips} and following sections. In Fig.~\ref{fig:EtaPlot1}, we plot $\eta(\omega)$ for the full SQUID chain with the \CMT{} RF amplifier, and in Fig.~\ref{fig:EpsilonPlot1}, we plot $\epsilon_{\rm uc}(\omega)$. 

%
%
%
%